\documentclass[twocolumn,tighten,usenames,dvipsnames,twocolappendix]{aastex631}
\usepackage[normalem]{ulem}
\usepackage{verbatim}
\usepackage{amssymb}
\usepackage{amsmath}
\usepackage{xspace}
\usepackage{graphicx}
\usepackage{lipsum}
\usepackage{multirow}
\usepackage{amsmath,mathtools,mathrsfs}
\usepackage{hyperref}
\usepackage{afterpage}
\usepackage{booktabs}
\usepackage{enumitem}

\definecolor{mygreen}{RGB}{150,150,10} 

\newcommand{\s}{{\rm s}}

\newcommand{\K}{{\rm K}}

\newcommand{\VirA}{M87*\xspace}
\newcommand{\SgrA}{Sgr~A*\xspace}

\defcitealias{M87PaperI}{M87*~Paper~I}
\defcitealias{M87PaperII}{M87*~Paper~II}
\defcitealias{M87PaperIII}{M87*~Paper~III}
\defcitealias{M87PaperIV}{M87*~Paper~IV}
\defcitealias{M87PaperV}{M87*~Paper~V}
\defcitealias{M87PaperVI}{M87*~Paper~VI}
\defcitealias{M87PaperVII}{M87*~Paper~VII}
\defcitealias{M87PaperVIII}{M87*~Paper~VIII}
\defcitealias{M87PaperIX}{M87*~Paper~IX}

\defcitealias{SgrAPaperI}{Sgr~A*~Paper~I}
\defcitealias{SgrAPaperII}{Sgr~A*~Paper~II}
\defcitealias{SgrAPaperIII}{Sgr~A*~Paper~III}
\defcitealias{SgrAPaperIV}{Sgr~A*~Paper~IV}
\defcitealias{SgrAPaperV}{Sgr~A*~Paper~V}
\defcitealias{SgrAPaperVI}{Sgr~A*~Paper~VI}
\defcitealias{SgrAPaperVII}{Sgr~A*~Paper~VII}
\defcitealias{SgrAPaperVIII}{Sgr~A*~Paper~VIII}

\shorttitle{Foiling Black Hole Foils}
\shortauthors{Faraji \& Broderick}

\begin{document}

\title{Foiling Black Hole Foils:
        Revealing Horizon Alternatives with Baryonic Atmospheres}


\author[0000-0003-1727-8992]{Shokoufe Faraji}
\affiliation{Department of Physics and Astronomy, University of Waterloo, 200 University Avenue West, Waterloo, ON, N2L 3G1, Canada}
\affiliation{Waterloo Centre for Astrophysics, University of Waterloo, Waterloo, ON N2L 3G1 Canada}
\affiliation{Perimeter Institute for Theoretical Physics, 31 Caroline Street North, Waterloo, ON, N2L 2Y5, Canada}

\author[0000-0002-3351-760X]{Avery E. Broderick}
\affiliation{Perimeter Institute for Theoretical Physics, 31 Caroline Street North, Waterloo, ON, N2L 2Y5, Canada}
\affiliation{Department of Physics and Astronomy, University of Waterloo, 200 University Avenue West, Waterloo, ON, N2L 3G1, Canada}
\affiliation{Waterloo Centre for Astrophysics, University of Waterloo, Waterloo, ON N2L 3G1 Canada}

 \begin{abstract}
Event horizons are a defining feature of black holes. Consequently, there have been many efforts to probe their existence in astrophysical black hole candidates, spanning ten orders of magnitude in mass. 
Nevertheless, horizons remain an obstacle to unifying general relativity and quantum mechanics, most sharply presented by the information paradox. 
This has motivated a proliferation of horizonless alternatives (black hole foils) that avoid event horizons and are therefore presumably benign.

We explore the fate of accreting material in the context of spherically symmetric horizonless compact objects in the steady state regime.
We find that for typical accreting astronomical targets, foils can develop an optically thick, convectively stable baryonic atmosphere, immediately implying the creation of an emitting photosphere at modest redshift even when the surface redshift is extreme.
The emergent photosphere luminosity is driven toward the accretion powered equilibrium value and is only weakly sensitive to the foil redshift. 
Moreover, local gas-surface interaction provides a microphysical lower bound on the effective base temperature, insulating the atmosphere from arbitrarily cold foils. 
The unknown properties of the foil enter only through local boundary conditions controlling baryon processing and thermal coupling at the surface, making the solutions broadly applicable to horizonless alternatives that do not invoke significant additional nonlocal interactions. 
Thus, under minimal assumptions (GR exterior and local surface interactions), spherically symmetric horizonless foils in accreting systems are generically observationally exposed: the absence of a thermal photosphere directly constrains or rules out broad classes of such models.
\end{abstract}

\section{Introduction}
\label{sec:intro}

With the advent of direct near-horizon probes, via gravitational waves \citep{LIGO_FirstDetection,LIGO_Kerr} or direct imaging \citep{M87_PaperI,SgrAPaperI}, the study of black hole horizons\footnote{The teleological definition of the event horizon renders it operationally subtle as an empirical object.
Nevertheless, Hawking-like particle creation does not require a global event horizon; a slowly evolving apparent/trapping horizon (or, more generally, an approximately exponential redshift relation between $\mathscr{I}^-$ and $\mathscr{I}^+$) is sufficient \citep{2003IJMPD..12..649V,2011JHEP...02..003B}.
therefore we focus our attention on what we might call ``astrophysical'' horizons, i.e., objects that look like black holes on astronomically relevant timescales.  It is to these that we shall refer to as horizons henceforth.}

has become an empirical effort. 
Nevertheless, black hole horizons remain a significant theoretical impediment to the long sought unification of general relativity and quantum mechanics, often described via the information paradox and its cousins 
\citep{Hawking1976,Preskill1992,Mathur2009,AMPS2013}.
One way that these conceptual concerns can be retired is to simply avoid the existence of black hole horizons altogether, i.e., do away with black holes and replace them with some object that is causally connected to the external universe \citep{Mathur2002Proposal,2024arXiv241209495M}.

While significant effort has been devoted to deriving horizon-scale modifications from candidate microscopic theories or controlled semiclassical frameworks -- including string-theoretic
microstate geometries/fuzzballs \citep{2005ForPh..53..793M,2008PhR...467..117S,2007hep.th....1216B,2019arXiv191213108W,2025arXiv250317310B}, horizonless ultracompact solutions in asymptotically free quadratic gravity \citep{2017PhRvD..95h4034H},
and semiclassical backreaction models \citep{PhysRevD.77.044032,2022NatSR..1215958A} -- explicit solutions that are
demonstrably generic, dynamically formed, and astrophysically complete remain limited.

In contrast, a broad class of black hole foils, typically qualitatively motivated alternatives that avoid an event horizon by introducing novel, often unknown, physics (e.g., gravastars \citep{MazurMottola2004,VisserWiltshire2004},
ultracompact fluid configurations tied to Buchdahl-type limits \citep{Buchdahl1959,DanielssonGiri_2021}, boson stars \citep{Kaup1968,RuffiniBonazzola1969,LieblingPalenzuela2012};
see also the review \citep{CardosoPani2019}).  
Because the information paradox would arise should a black hole horizon exist anywhere in the universe, such programs must imagine replacing the entire mass range of observed black hole candidates, from roughly $5\,M_\odot$ to $7\times10^{9}\,M_\odot$ \citep{LIGO_SmallBH,M87_PaperVI}. 
Therefore, even absent an underlying fundamental theory, black hole foils are useful parameterizations of the observational constraints on the existence of black hole horizons, hence their name.

Over the past quarter century, a class of arguments for the observational existence of black hole horizons based on accreting black hole candidate systems has been proposed 
\citep{NarayanGarciaMcClintock:1997,NarayanHeyl:2002,McClintockNarayanRybicki:2004,BN07_gravastars,NarayanMcClintock:2008,BLN09_sgra,BNK15_m87}.
Essentially, these invoke the possibility of an observable thermal photosphere powered by the kinetic energy within the accretion flow as it plunges onto the black hole foil. In steady state, the 
luminosity 
measured at infinity from a thermal photosphere,
\begin{equation}
    A_\infty \sigma T_\infty^4 \approx \eta \dot{M} c^2,
\end{equation}
is balanced by the incoming accretion power, where $A_\infty$, $T_\infty$, and $\dot{M}$ are the apparent photosphere area, temperature, and mass accretion rate as measured by an observer at infinity, $\sigma$ is the Stefan-Boltzmann constant, and $\eta\sim0.5$ is an efficiency factor describing the kinetic energy still held by the accreting matter as it plunges into the photosphere.
Very compact surfaces, i.e., black hole foils with high surface redshifts, simplify the argument via the strong gravitational lensing that serves to couple the surface to itself, facilitating thermalization.
The presence of such photospheres has been excluded at overwhelming confidence in the two horizon-scale Event Horizon Telescope targets, \VirA and \SgrA \citep{M87_PaperVI,SgrAPaperVI,Broderick2023}.
The remaining debate centers on when and if a given black hole foil might develop a thermally emitting surface.

Unfortunately, due to their poorly defined nature, the physical properties of black hole foil surfaces are inherently difficult to generally assess.
For example, some foils invoke near-infinite heat capacities and near-zero surface temperatures, and thus seek to evade the observational constraints by never reaching steady state between the thermal surface luminosity and kinetic power of the accreting matter.\footnote{It is of interest that wherein a suitable microphysical model is provided, it may be possible to exclude these, e.g., \citet{BN07_gravastars}.  It would seem, therefore, that this lack of definition may be a ``feature'', though one that should be challenged.}
However, these complications could be avoided if the photosphere was a feature of the accretion flow itself, which is comprised of well understood baryonic matter and electromagnetic fields.  
Such a simplification was instrumental in recent constraints on the existence of naked singularities  
\citep{VieiraWlodek_2023,2024ApJ...977..249B}. 
Where there it was motivated by the particular gravitational properties of curvature singularities, here it is a natural consequence of the collection and relaxation of the accreting matter within the boundary layer of the settling flow.

In constructing a suitable baryonic atmosphere around arbitrary black hole foils, we make a pair of key assumptions:
\begin{enumerate}
\item Gravity may be well described outside of the foil surface by general relativity.
\item Any novel interactions between the accreting baryonic matter and the surface must be (nearly) local, i.e., they must occur within an appropriate ``small'' distance of the surface.
\end{enumerate}
The former appears well established, at least to order unity, by gravitational wave observations \citep{LIGO_AltGrav} and direct imaging \citep{Kocherlakota:2021,SgrAPaperVI,Broderick2023}.  
The latter may very well not be true, but deviations from locality are capable of solving the information paradox without the need for foils altogether, and thus form a separate class of solution.\footnote{While it certainly remains possible that physics deviates simultaneously in multiple ways from the standard picture near black hole candidates, we view this as a doubly extraordinary solution, and focus on the more ordinary singly extraordinary possibilities.}
We also adopt spherical symmetry for computational simplicity, but do not anticipate this to be a significant qualitative restriction.
The properties of the black hole foil are distilled into two local pieces of boundary data: one controlling the inward baryon flux (a mass transmittance parameter) and one controlling the exchange of heat at the surface (an energy exchange condition expressed through the base temperature).
  
In this way, we construct a widely applicable set of astrophysical models of a well-behaved thermal photospheres that envelop and thereby reveal black hole foils, independent of the specific underlying physical mechanisms responsible for avoiding the appearance of a horizon.

In this paper the term "foil" is used in a deliberately operational sense: it denotes any long-lived, horizonless compact object that presents an interface at an areal radius, $r_f$, to which accreting baryons can remain causally connected. We do not assume a specific microscopic theory or interior equation of state, and we do not attempt to construct a global solution of the field equations for the object as well. Instead, we treat the foil as a timelike inner boundary for the exterior accretion/atmosphere problem: the spacetime for $r\ge r_f$ is taken to be well-described by the Schwarzschild geometry, while the unknown microphysics at $r=r_f$, enters only through boundary data controlling (i) the exchange of heat and (ii) the steady baryon flux into/through the surface. This formulation is therefore agnostic as to whether the interior completion arises from exotic matter in general relativity, a modified gravity theory, or a quantum gravity completion, provided the exterior metric is close to Schwarzschild and the interaction with baryons is approximately local.

Specifically, we construct a minimal relativistic model of a steady, spherically symmetric, subsonic baryonic settling layer (atmosphere) between an outer photosphere and a physical surface (foil) supported by internal pressure gradients. 
Energy is transported through the atmosphere by radiative diffusion, while mass is carried by slow subsonic settling, subject to boundary conditions at an outer accretion shock (where power and momentum are supplied) and at the foil surface.

We also emphasize that we do not address if such horizonless configurations
form generically as the end states of gravitational collapse, nor do we model their dynamical stability. Our results are conditional in the following sense:
if a long-lived horizonless compact object with an exterior close to
Schwarzschild exists and is fed by sustained accretion for long times compared to the settling and diffusion timescales, then a quasi-stationary baryonic atmosphere forms whose structure and emergent luminosity can be analyzed using the conservation laws and diffusion closure developed in this paper.

We explicitly derive the radiative diffusion law (and thus conservation of luminosity as seen by observers at infinity) from a stress-energy conservation and the associated timelike Killing current.  
In appendices, we  discuss the physical origins of opacity within the atmosphere and explore a variety of modifications to the simplest picture, incorporating radiation pressure and inertia, various radiative transfer processes, relaxing the slow-settling condition and validity of the steady approximation; we find that in all cases the photosphere luminosity remains close to the steady state, accretion powered value for optically thick atmospheres.

Therefore, this framework provides a controlled way to assess whether redshift and diffusion can hide accretion power in surface atmospheres and thus places sharp, transparent constraints on horizonless alternatives. The paper is organized as follows: Section \ref{sec:eqoas} states equations of atmospheric structure and assumptions. Section \ref{sec:nonrelativistic} presents the nonrelativistic equations for intuition,
followed by the full relativistic system, its rationalized formulation, and approximate solutions in the compact surface limit in Section \ref{sec:relativistic}. We then state the boundary/viability conditions, microphysical base condition and the resulting bounds in Section \ref{sec:boundary}. Section \ref{sec:conclusions} summarizes the results and conclusions.


 \section{ Atmospheric Structure} \label{sec:eqoas}

We consider a horizonless compact object (a foil) with a material surface at radius $r_f$, whose exterior spacetime is well described by the Schwarzschild solution\footnote{The foil is not modeled as a complete interior solution; it is a boundary at $r=r_f$ supplying the inner data required to close the steady first order structure equations in the exterior. All conclusions therefore rely on to the exterior settling layer and do not assume a particular interior stress energy, only that the exterior is approximately Schwarzschild and that the surface interaction
with baryons is local on scales $\ll H$.}. Because accreting baryons do not disappear through a horizon, continuous supply naturally leads to the formation of a quasi-stationary boundary layer: a pressure-supported settling atmosphere extending from an outer photosphere (either an accretion shock or a smooth matching surface to the inflow) down to the foil. We adopt a minimal description in which the atmosphere is steady, spherically symmetric, and subsonic, with gas pressure providing the dominant support, and with radiative energy transport captured by diffusion through a layer that is geometrically thin yet optically thick. Under these assumptions the structure is determined by radial transport equations for mass, momentum, and energy, closed by boundary conditions at the outer photosphere (where the inflow supplies power and momentum) and at the foil surface (where the microphysics controls the exchange of mass and heat). In steady, spherically symmetric flow the interior solution is determined by first order radial transport equations, so inner data must be specified at the foil to fix how mass and heat are exchanged; otherwise the problem is underdetermined and the interior cannot be matched uniquely to the outer photosphere or shock. We therefore permit an extreme range of possibilities by setting:

\begin{enumerate}
\item Surface temperature $T_f$.  This may differ substantially from the temperature of the accreting gas. For example, in gravastar models with very large heat capacity and even nearly infinite, one might have $T_f \approx 0~{\rm K}$.

\item Inward mass flux at the surface $J_f$. We adopt a linear closure with the density just above the surface,
\begin{equation}\label{eq:surface-flux}
J_f \,=\, r_f^2\,\mathcal{T}\,\rho_f\,v_{\rm th},
\end{equation}
where $r_f$ is the surface radius, $\rho_f$ the baryon density immediately above the surface, $v_{\rm th}$ a characteristic thermal speed, and $\mathcal{T}$ a dimensionless surface transmission coefficient that encodes the microphysics of baryon absorption by (or conversion at) the foil.
Such a mechanical picture turns out to be practically convenient.
\end{enumerate}
Here and throughout, $\rho_f \equiv \rho(r_f^+)$ and $T_f \equiv T(r_f^+)$ denote the baryon density and gas temperature immediately outside the foil surface.
The second condition allows the foil either to absorb the accreting baryons or to convert them into other forms of matter.  We choose the sign convention that $J_f>0$ denotes inward mass flow. The form \autoref{eq:surface-flux} matches the conserved spherical mass flux used below: in steady flow $J$ is constant with radius and equals $J_f$ at the surface, with the convention that $J>0$ denotes inward accretion. All steady solutions considered below assume $0<\mathcal T<1$. For nonzero external supply, the endpoint $\mathcal T=0$ admits no finite density stationary transmitted mass flux under the present boundary condition; matter then accumulates unless an additional outflow or conversion channel is specified. We therefore treat this endpoint as a time dependent problem rather than as a member of the steady solution family.\footnote{$[J_f] = MT^{-1}$; indeed, $[r_f^2\rho_f v_{\rm th}]=L^2\times M L^{-3}\times L T^{-1}=M T^{-1}$. Angular factors and distribution function constants are absorbed into $\mathcal{T}$.}

We will also make the following simplifying assumptions, none of which we expect to be violated in practice, and many of which we are able to subsequently verify:

\begin{enumerate}
\item Steady state. Therefore we have $\nabla_\mu(\rho u^\mu)=0$ and the Killing energy flux is conserved ($dL_\infty/dr=0$). A posteriori, the external supply must persist approximately unchanged for longer than the layer formation and response time, ($t_{\rm sup}\gg t_{\rm form}\equiv\max(t_{\rm set},t_{\rm diff}).$) Departures from steady state and the associated energy-storage relation are discussed in \autoref{app:timedep}.

\item Spherical symmetry. Angular structure is neglected so all quantities depend only on $r$. Any residual rotation, magnetic stresses, or multipoles are assumed small enough to enter as higher-order corrections to the radial balance.

\item Slow (subsonic) settling. $v^r\ll c$ and $M\equiv |v^r|/c_s\ll1$ with $c_s^2=kT/m$. This ensures pressure support dominates, no sonic point forms in the atmospheric layer, and is consistent with the surface transmission bound $|\mathcal{T}|<1$ used below.

\item Photon pressure subdominant. The gas is ideal and baryon–dominated: $P_{\rm rad}\ll P_{\rm gas}$ with
\begin{equation}
 \frac{P_{\rm rad}}{P_{\rm gas}}
  = \frac{(1/3)\,aT^4}{\rho kT/m}
 = \frac{a m}{3k}\,\frac{T^{3}}{\rho} \,\ll\,1,
\end{equation}
where $a=4\sigma/c$. We will monitor this ratio in the solutions to validate the closure.

\item Geometrically thin atmosphere. The scale height
\begin{equation}\label{eq:Hassump}
H \,\equiv\, \frac{kT}{m|g|},\qquad
|g(r)|=\frac{GM}{N(r)\,r^{2}}.
\end{equation}
The pressure scale height satisfies $H/r_f\ll 1$ (equivalently $\Delta r\sim N_f H\ll r_f$). This permits evaluating geometric prefactors at $r_f$ (e.g.\ $r\simeq r_f$ outside derivatives) while retaining the exact lapse $N(r)$, whose variation can remain $O(1)$ in the compact surface limit.

\item Exterior Schwarzschild geometry. Outside the foil the spacetime is static and spherically symmetric,
\begin{equation}\label{eq:METRIC}
ds^2=-N^2(r)\,dt^2+N^{-2}(r)\,dr^2+r^2 d\Omega^2,
\end{equation}
where
\begin{equation}\label{eq:N-METRIC}
N^2(r)=1-\frac{2GM}{rc^2}.
\end{equation}
This is appropriate when rotation and frame dragging are negligible for the atmospheric layer; any such effects would appear as controlled corrections.

Note that in the Schwarzschild exterior the foil radius $r_f$ is equivalently characterized by the lapse $N_f \equiv N(r_f)=\sqrt{1-2r_g/r_f}$ and the gravitational redshift of photons emitted from a static surface, $z_f = N_f^{-1}-1$.
 In a concrete horizonless model $N_f$ is not free: it is fixed by the interior field equations and the microphysics that replaces the horizon. In this paper, we intentionally allow such horizonless "foil" constructions and ask what the surrounding baryonic atmosphere implies observationally once the exterior geometry is Schwarzschild.


In fact, in general relativity, not all values of $N_f$ are realizable by ordinary matter.
For static, spherically symmetric isotropic perfect fluid configurations satisfying
Buchdahl's assumptions one has $2GM/(r_f c^2)\le 8/9$, implying
$N_f\ge 1/3$ (equivalently $z_f\le 2$) \cite{Buchdahl1959,Lindblom1984}.
Achieving $N_f\ll 1$ therefore requires violating at least one of these assumptions,
e.g. through anisotropic stresses or thin shell constructions (as in gravastar type
models) \cite{VisserWiltshire2004}.

Realistic equations of state for neutron stars typically impose even tighter redshift bounds. Conversely, constructions that relax the Buchdahl assumptions via strong anisotropy or thin shells (e.g.\ gravastars) can in principle reach $r_f\to 2r_g$ and hence $N_f\ll 1$. If significant rotation is present, sufficiently compact horizonless objects can develop an ergoregion, which can drive an instability and thereby constrain extremely small $N_f$ in long-lived astrophysical systems \cite{Cardoso2008}.

In this work we therefore treat $N_f$ (or equivalently $r_f/r_g$) as an external parameter to keep the atmospheric calculation model agnostic; the analysis below shows how the atmosphere maps a given $N_f$ to an emitting photosphere at $r_s$ with redshift $N_s$ under the stated transport and boundary assumptions. Finally, while varying $N_f$ can in principle change the apparent area $A_\infty\propto N^{-2}$ for a photosphere outside the photon sphere, and the mapping between local and redshifted temperatures, our main result is that for an optically thick, diffusion dominated settling layer the emergent luminosity is driven close to the accretion power and is only weakly sensitive to $N_f$ (Sections~\ref{sec:relativistic}-\ref{sec:boundary}).

\item  Optical thick. The radiative transport law used throughout is the diffusion limit and therefore requires that the atmosphere be optically thick on the local pressure support scale (more discussion in \autoref{sec:thick}.) 
\end{enumerate}
Throughout, $T(r)\equiv T_{\rm loc}(r)$ denotes the local thermodynamic temperature measured in the comoving frame, and
\begin{equation}
T_\infty(r)\equiv N(r)\,T(r)
\end{equation}
denotes the corresponding redshifted temperature. We use $T_{\rm eff}$ only for the flux defined temperature of the emitting photosphere, $F_{\rm loc}=\sigma T_{\rm eff}^{4}$. With the blackbody photosphere condition adopted here, $T_{\rm eff}=T_s$ and
$T_{{\rm eff},\infty}=N_s\,T_{\rm eff}=T_{\infty,s}$. Away from the photosphere, $T_\infty$ denotes the redshifted local gas temperature and should not be interpreted as an effective temperature.


\section{Non-relativistic Limit}\label{sec:nonrelativistic}

We begin with the relevant equations in the non-relativistic limit, both as a comparison for the subsequent fully relativistic treatment and as a tool for building intuition about the resulting solution. The equations we utilize are the standard (nonrelativistic) stellar-structure equations specialized to the case
without volumetric heating ($dL/dr=0$)
(see, e.g., \cite{1994sse..book.....K}). Specifically, these are continuity
\begin{equation}    \label{eq:NR_continuity}
\frac{d}{dr}\,\left(r^2 \rho v^r\right) = 0,
\end{equation}
the radial component of the Euler equation
\begin{equation}\label{eq:NR_Euler}
\rho v^r \frac{d}{dr} v^r = -\frac{d}{dr} P + \rho g,
\end{equation}
where $P=\rho k T/m$ (ideal gas; $m$ is the mean particle mass) and $g$ is the gravitational acceleration, and the radiative diffusion equation
\begin{equation}\label{eq:NR_RadDiff}
\frac{d T}{dr} = - \frac{3\,\kappa\,\rho\,L}{64\,\pi\,\sigma\,r^2\,T^3},
\end{equation}
where $\sigma$ is the Stefan-Boltzmann constant, $L$ is the outward luminosity carried through a sphere of radius $r$, and $\kappa$ is the (here taken constant) opacity, dominated by electron scattering \footnote{\autoref{eq:NR_RadDiff} is the standard form $ \tfrac{d}{dr} T = -(3\kappa\rho L)/(16\pi a c\, r^2 T^3)$ rewritten using $a c = 4\sigma$.}.

We do not supplement this with an energy generation equation ($dL/dr=4\pi r^2 \rho \epsilon$), since no volumetric heating is expected; therefore $L$ is radially constant and is fixed by the boundary condition at the photosphere/shock. At the accretion shock or matching photosphere located at $r_s$, we take
\begin{equation}\label{eq:NR_LuminosityBC}
L = 4\pi r_s^2 \,\sigma T_s^4 - \eta \,\dot{M} \,c^2,
\end{equation}
with the first term the outward blackbody surface luminosity and the second the inward accretion power (so typically $L<0$). With this sign convention a negative $L$ implies $\partial_r T>0$ by  \autoref{eq:NR_RadDiff}; i.e., the temperature rises outward across the atmosphere, rendering the layer convectively stable (via the Schwarzschild criterion, CITE). The observed surface luminosity is the outward term, $L_{\rm surf}=4\pi r_s^2 \sigma T_s^4$, while the net energy flux through the atmosphere is $L$ (usually negative in the accretion-powered case).

In addition, continuity implies that $J \equiv r^2\rho v^r$ is constant and simultaneously set by the conditions at the accretion shock and the foil surface,
\begin{equation}
J=\frac{\dot{M}}{4\pi}= r_f^2\,\mathcal{T}\,\rho_f\,\sqrt{\frac{kT_f}{m}}.
\end{equation}
Thus we have a boundary condition on the atmosphere density,
\begin{equation}
\rho_f = \frac{\dot{M}}{4\pi r_f^2 \mathcal{T} \sqrt{kT_f/m}} .
\end{equation}
The radial Euler equation may be reconfigured to remove $v^r=J/(r^2\rho)$.  Using the thin layer approximation $r\simeq r_f$ (so $r$ dependences outside derivatives are evaluated at $r_f$), and writing $dP/dr=(k/m)\,(T\,d\rho/dr+\rho\,dT/dr)$ and substituting
$v^r=J/(r_f^2\rho)$ into $\rho v^r dv^r/dr$; then inserting the diffusion law
\autoref{eq:NR_RadDiff},
yields
\begin{equation}  \label{eq:NR_rho}
\frac{d\rho}{dr}
\,\approx\,
\rho\, \frac{\displaystyle g \,+\, \frac{3k\,\kappa\,\rho\,L}{64\pi\, r_f^2\, m\, \sigma\, T^3}}{\displaystyle \frac{kT}{m} \,-\, \frac{J^2}{r_f^4\,\rho^2}} \, .
\end{equation}
The outward luminosity $L$ is negative in the accretion powered case, and $g=-GM/r^2<0$, so both terms in the numerator are negative. The denominator in \autoref{eq:NR_rho} is the usual sonic discriminant: it approximately reduces to $c_s^2-(v^r)^2$ with $c_s^2=kT/m$. Because the settling flow is subsonic by assumption, the denominator is positive. Therefore $\rho(r)$ decreases monotonically with radius and, correspondingly, $v^r$ increases monotonically.

This condition places a natural constraint on $\mathcal{T}$ at $r_f$:
\begin{equation}
\frac{kT}{m} - \frac{J^2}{r_f^4\rho_f^2}=  \frac{kT_f}{m}\,(1 - \mathcal{T}^2) > 0
\,\,\Rightarrow\,\,|\mathcal{T}|<1.
\end{equation}
Combined, \autoref{eq:NR_RadDiff} and \autoref{eq:NR_rho} define an initial value problem given $(\rho_f,T_f)$; the equations may be integrated outward until the outer boundary condition is satisfied. While \autoref{eq:NR_LuminosityBC} fixes the net luminosity, we still require the shock (photosphere) radius $r_s$. The latter is set by the jump conditions; a useful estimate is
\begin{equation}\label{eq:PS}
P_s \,\approx\, \frac{\dot{M}\,c}{4\pi\,r_s^2},
\end{equation}
i.e., the post shock pressure balances the ram pressure $\rho_a u_a^2$ with
$4\pi r_s^2\rho_a u_a=\dot M$ and $u_a\sim c$ near the shock.


\subsection{Characteristics of Solutions}
Even without strong field corrections, the nonrelativistic (NR) limit already fixes the qualitative architecture of any steady, pressure supported settling layer. It yields two robust predictions that do not depend on microphysics: (i) the sonic discriminant at the base must remain positive (no hidden sonic point), and (ii) when the net luminosity $L\le 0$ the pressure profile is strictly decreasing. Both statements survive verbatim in the relativistic treatment once the redshift factors $N(r)$ are restored; only coefficients change. These NR results therefore provide clean, model independent diagnostics that we will carry forward.

First, if the denominator of \autoref{eq:NR_rho} is negative, the density gradient becomes positive and the configuration loses pressure support at the base (the flow there is supersonic). We therefore require
\begin{equation}
    \sqrt{\frac{kT_f}{m}} \,>\, v^r_f \,=\, \frac{J}{r_f^2\,\rho_f}.
\end{equation}
Second, the pressure gradient is generically negative throughout the subsonic settling layer. Starting from Euler,
\begin{equation}
\frac{dP}{dr} \,=\, \rho\,g \,+\, (v^r)^2\,\frac{d\rho}{dr},
\qquad v^r=\frac{J}{r_f^2\rho},
\end{equation}
and substituting \autoref{eq:NR_rho} (with the thin layer approximation $r\simeq r_f$) gives
\begin{equation}\label{eq:NR_dP_clean}
\frac{dP}{dr}\,=\,
\frac{\displaystyle \rho\,\frac{kT}{m}\,g \;+\;
\displaystyle \frac{3k\,\kappa\,L}{64\pi\,\sigma\, r_f^2\, m\, T^3}\,\frac{J^2}{r_f^4}}
{\displaystyle \frac{kT}{m} \,-\, \frac{J^2}{r_f^4\,\rho^2}} \, .
\end{equation}
The substitution cancels only the $g(v^r)^2$ piece generated by the inertial term, leaving the expected hydrostatic contribution (and recovering $dP/dr\simeq \rho g$ when $(v^r)^2\ll kT/m$). For accretion-powered cases $L\le 0$ and $g<0$, while the denominator is positive in the subsonic regime, so $dP/dr<0$ throughout the settling layer.

A direct consequence is that the pressure at the base must exceed the post shock pressure; otherwise the outer jump conditions cannot be satisfied. Using the standard ram‑pressure estimate at the shock \autoref{eq:PS}, we require
\begin{equation}
P_f \,>\, P_s
\,\,\Longrightarrow\,\,
\rho_f\,\frac{kT_f}{m} \,>\, \frac{\dot{M}\,c}{4\pi\,r_s^2}\,\simeq\, \frac{\dot{M}\,c}{4\pi\,r_f^2},
\end{equation}
where in the last step we used $r_s\simeq r_f$ (Order unity factors in the shock jump can be absorbed in this estimate; the inequality is the key requirement.).

\subsection{Approximate Solution}
To expose the dominant scalings, we now construct an approximate solution in a regime where the temperature varies only weakly across one scale height and the base flow is strongly subsonic. Concretely, we assume:
\begin{enumerate}
\item Weak temperature variation across $H$. This is the precise condition that allows us to treat $T$ as approximately constant when solving the hydrostatic balance.
\item Strong pressure support at the base. The flow is strongly subsonic there:
$c_s \gg v^r$, i.e.,
.
$kT_f/m \gg J^2/(r_f^4 \rho^2)$.
\end{enumerate}

Density profile over the height scale: the consequence of the first one is temperature varies only weakly across a scale height ($T$ nearly constant), while the density adjusts rapidly. Taking $T\simeq \bar T$ constant in the hydrostatic term,
\begin{equation}
\frac{dP}{dr}=\rho g,\qquad P=\frac{\rho k\bar T}{m},
\end{equation}
gives
\begin{equation}
\frac{d\rho}{dr} \simeq \rho \frac{m g}{k\bar T}
\,\,\Rightarrow\,\,
\rho(r)\simeq \rho_f\,\exp\,\left[-\frac{(r-r_f)}{H}\right],
\end{equation}
where  $H$ is given by \autoref{eq:Hassump}. Thus, the density falls exponentially over the height scale $H$.

\medskip
Temperature profile over the height scale: Inserting $\rho\simeq \bar\rho$ (a layer average) into the diffusion equation, taking $r\simeq r_f$, multiplying by $4T^3$ and integrating across one scale height gives
\begin{equation}
T_s^{4}\,\simeq\, T_f^{4}\left[1-\frac{3\,\bar\tau\,L}{16\,\pi\,r_f^{2}\,\sigma\,T_f^{4}}\right],
\label{eq:Ts4-approx}
\end{equation}
where $\bar\tau \,\equiv\, \kappa\,\bar\rho\,H$ is the optical depth across one scale height. Because accretion power drives $L\le 0$ (outward positive), the bracket exceeds unity and $T$ rises outward consistent with unconditional convective stability.

\begin{figure}
    \centering
    \includegraphics[width=\columnwidth]{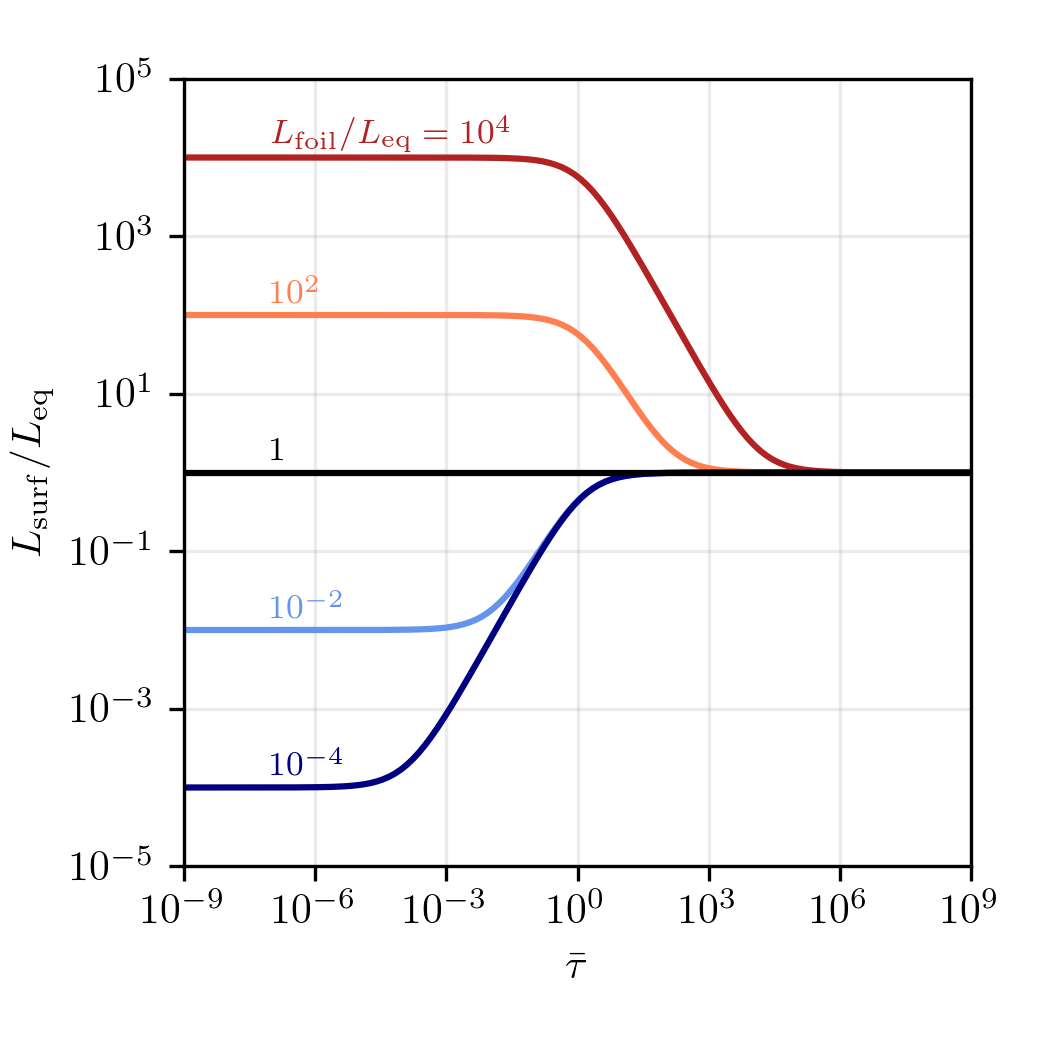}
    \caption{Surface luminosity at the top of the baryonic atmosphere as a function of atmosphere optical depth across one scale height for various foil luminosities. Curves correspond to different intrinsic foil luminosities $L_{\rm foil}/L_{\rm eq}$ (labels). The thin limit behavior is an extrapolation. In the optically thin limit $\bar\tau\ll 1$, the layer is transparent and $L_{\rm surf}\simeq L_{\rm foil}$. In the optically thick limit $(3/4)\bar\tau\gg 1$, diffusion forces the emergent luminosity to $L_{\rm surf}\to L_{\rm eq}$, essentially independent of $L_{\rm foil}$.}
    \label{fig:Lvtau}
\end{figure}

\medskip
Surface luminosity: with $L_{\rm surf}\equiv 4\pi r_s^2 \sigma T_s^4\simeq 4\pi r_f^2\sigma T_s^4$ and $L=L_{\rm surf}-\eta\dot M c^2$, \autoref{eq:Ts4-approx} yields
\begin{equation}\label{eq:Lsurf-approx}
L_{\rm surf}
\,\simeq\,
\frac{\,L_{\rm foil} + \displaystyle \frac{3}{4}\,\bar\tau\,\eta\,\dot M c^2\,}{1+\displaystyle \frac{3}{4}\,\bar\tau},
\end{equation}
where $L_{\rm foil}\equiv 4\pi r_f^2 \sigma T_f^4$. \autoref{eq:Lsurf-approx} shows that the photospheric luminosity is controlled by a single dimensionless parameter, $(3/4)\,\bar{\tau}$, This parameter characterizes how effectively radiation is retained and reprocessed within one
pressure scale height before escaping. In the strongly optically thick regime, $(3/4)\,\bar{\tau}\gg 1$, the surface luminosity is driven to the accretion-powered equilibrium value $L_{\mathrm{surf}}\to \eta \dot{M}c^{2}\equiv L_{\mathrm{eq}}$, essentially independent of the foil temperature. In the opposite regime, $(3/4)\,\bar{\tau}\ll 1$, the atmosphere cannot efficiently communicate the accretion power to the photosphere and the emergent luminosity relaxes toward the intrinsic foil luminosity $L_{\mathrm{surf}}\to L_{\mathrm{foil}}$. \autoref{fig:Lvtau} illustrates this interpolation for several values of $L_{\mathrm{foil}}/L_{\mathrm{eq}}$.

\medskip
\noindent 

Using the surface boundary condition
\begin{equation}
\rho_f=\frac{\dot M}{4\pi r_f^2\,\mathcal{T}\,v_{\rm th,f}},
\end{equation}
given
\begin{align}
&v_{\rm th,f}\equiv\sqrt{kT_f/m}\,,
\end{align}
and $H$ is given by \autoref{eq:Hassump}, we obtain
\begin{equation}\label{eq:tau-estimate}
\bar\tau \,\approx\,
\frac{\kappa\,\dot M}{4\pi\,\mathcal{T}\,GM}\,
\frac{k\bar T/m}{v_{\rm th,f}}
\,=\,
\frac{\kappa\,\dot M}{4\pi\,\mathcal{T}\,GM}\,
\frac{c_s^2(\bar T)}{v_{\rm th,f}}\,.
\end{equation}
The approximation is self-consistent when

\begin{equation}
\varepsilon_T \,\equiv\, \frac{3\,\bar\tau\,|L|}{16\,\pi\,r_f^{2}\,\sigma\,T_f^{4}}
\,\ll\,1 ,
\end{equation}
i.e., when the fractional change in $T^4$ across one scale height is small. In this regime the temperature is nearly constant over $H$ and the optical depth $\bar\tau\gtrsim 1$, so our use of a single characteristic
temperature and density within the layer is justified. Equations \eqref{eq:Lsurf-approx}-\eqref{eq:tau-estimate} then quantify precisely when the surface radiates at $\eta\dot M c^2$ and when it relaxes toward $L_{\rm foil}$ for $(3/4)\bar{\tau}\ll1$; this is the small $\bar{\tau}$ limit of this diffusion based estimate. In particular, for fixed $\bar\tau$ the condition $\varepsilon_T\ll1$ corresponds to the near equilibrium branch in which the net luminosity through the layer satisfies $|L|\ll L_{\rm eq}$.

\subsection{Rationalized Form}
For explicit computations and to facilitate direct comparison with the relativistic case, we present a rationalized version of the equations of atmospheric structure that emphasizes the characteristic aspects of the above solutions. We define
\begin{subequations}
\begin{align}
\bar{x} &= \rho/\rho_f, \label{xbarNR} \\
\bar{y} &= T/T_{\rm eq}, \\
s &= (r-r_f)/h,
\end{align}
\end{subequations}
where the NR scale height at the equilibrium temperature is 
\begin{equation}\label{eq:h-equib}
  h\,\equiv  \frac{kT_{\rm eq}}{m|g|},
\end{equation}
and
\begin{equation}
L_{\rm eq}\,\equiv\,4\pi r_s^2 \sigma T_{\rm eq}^4 \,=\, \eta\,\dot{M}c^2.
\end{equation}
we take $r_s\simeq r_f$ within the thin layer. We also introduce the optical depth per height scale parameter 

\begin{equation}\label{eq:tau_h}
\tau_h \,\equiv\, \kappa\,\rho_f\,h .
\end{equation}
Using the thin layer approximation $r\simeq r_f$ in the coefficients, the dimensionless equations become
\begin{subequations}
\begin{align}
\partial_s \bar{x}&= \bar{x}\,
\frac{\displaystyle -1 \,+\, \frac{3\,\tau_h\,L}{16\,L_{\rm eq}}\,
\frac{\bar{x}}{\bar{y}^3}}{\displaystyle \bar{y} \,-\, \mathcal{T}^2\,
\frac{\bar{y}_f}{\bar{x}^2}},\\
\partial_s \bar{y}&= -\,\frac{3\,\tau_h\,L}{16\,L_{\rm eq}}\,
\frac{\bar{x}}{\bar{y}^3}\,,
\end{align}
\label{eq:NR_rEAS}
\end{subequations}
where the $-1$ in the numerator of $\partial_s\bar{x}$ is the (dimensionless) hydrostatic term $hg/(kT_{\rm eq}/m)=-1$, and the coefficient follows from the NR diffusion law with $ac=4\sigma$ (hence the $16$).

The boundary conditions are
\begin{subequations}
\begin{align}
&\bar{x}_f = 1,\\
&\bar{y}_f = T_f/T_{\rm eq},\\
&\bar{x}_s\,\bar{y}_s
= \frac{m}{\rho_f k T_{\rm eq}}\,
\frac{\dot{M}\,c}{4\pi r_s^2}
\,=\, \left(\frac{r_f}{r_s}\right)^{2}\frac{\mathcal{T}\,c}{\sqrt{|g|\,h}}\,
\bar{y}_f^{1/2},\\
&L = L_{\rm eq}\,\big( \bar{y}_s^4 - 1 \big),
\end{align}
\label{eq:NR_rBCs}
\end{subequations}
where we used $J=\dot M/(4\pi)=r_f^2\mathcal{T}\rho_f\sqrt{kT_f/m}$ in the second equality of the third line. The shock/outer pressure condition enters via
$\rho_s(kT_s/m)\simeq \dot{M}c/(4\pi r_s^2)$. Further, $L$ is radially constant in the layer, fixed by the outer boundary condition.


\section{Relativistic Limit} \label{sec:relativistic}

Because the atmospheric layer lies at high redshift ($N\ll 1$), and therefore relativistic effects control the dynamics. Nevertheless, we pursue an analogous path as in the NR derivation but work with the proper variables $(\rho,u^\mu)$ and the redshift invariant observables $(T_\infty, L_\infty)$, so the resulting equations retain the familiar structure with explicit $N(r)$ factors.

\subsection{Continuity equation}
We begin with the continuity equation for the rest-mass (baryon) current:
\begin{equation}
\nabla_\mu(\rho\,u^\mu)
\,=\,
\frac{1}{\sqrt{-g}}\,
\partial_\mu\,\left(\sqrt{-g}\,\rho\,u^\mu\right)
\,=\,0,
\end{equation}
where $\rho$ is the proper density, $u^\mu$ the fluid four-velocity, and $g$ the metric determinant.

Assuming stationarity ($\partial_t=0$), spherical symmetry, and purely radial motion
$u^\mu=(u^t,u^r,0,0)$, continuity reduces to
\begin{equation}
\partial_r\,\big(r^2\,\rho\,u^r\big)=0
\,\,\Longrightarrow\,\,
r^2\,\rho\,u^r=\text{const}.
\label{eq:R_continuity}
\end{equation}
This is the relativistic analogue of \autoref{eq:NR_continuity}; the invariance arises because $\sqrt{-g}=r^2\sin\theta$ in Schwarzschild. Inward flow has $u^r<0$ (since $r$ increases outward). To keep the inward mass flux positive, we define
\begin{equation}\label{eq:mass_flux}
J \,\equiv\, -\,r^2\,\rho\,u^r \,=\, \frac{\dot M}{4\pi} \,>\,0,
\end{equation}
which coincides with the non-relativistic definition used earlier (up to the sign choice).

\subsection{Radial Euler}
The Euler equation follows from local energy-momentum conservation for a perfect fluid,
\begin{equation}
T^{\mu\nu}=(\varepsilon+P)\,u^\mu u^\nu + P\,g^{\mu\nu},
\end{equation}
with proper energy density $\varepsilon$ and pressure $P$.

By contracting
$g_{r\mu}\nabla_\nu T^{\mu\nu}=0$ (equivalently $\nabla_\nu T_{r}{}^{\nu}=0$) and using spherical symmetry, after some algebra we obtain the steady radial Euler equation
\begin{align}\label{eq:R_Euler_full}
(\varepsilon+P)\,u^{r}\partial_r u_{r}
= &-\,(1+u_{r}u^{r})\,\partial_r P\\
&-(\varepsilon+P)\,\frac{N'}{N}\big(1+2\,u_{r}u^{r}\big).  \nonumber
\end{align}
In the regimes of interest, subsonic and low enthalpy regime \footnote{Here $\rho$ is the proper rest mass density used in the continuity equation. With $\varepsilon\simeq \rho c^2$ the distinction between $\varepsilon$ and $\rho$ is immaterial at our working order; factors of $c$ are kept explicitly in $N(r)$.}, the internal energy is small compared to the rest mass energy, ($|u_{r}u^{r}|\ll 1$, $P\ll\varepsilon$). Then \autoref{eq:R_Euler_full} reduces to

\begin{equation}
\rho\,u^r\,\partial_r u_r
= -\,\partial_r P - \frac{\rho\,{c^2}}{2}\,\partial_r\,\ln(N^2),
\label{eq:R_Euler}
\end{equation}
using $\varepsilon\simeq\rho c^2$ at our working order (rest-mass dominated gas; $P\ll \varepsilon$ and $|u_ru^r|\ll 1$). Indeed, in the static limit $u^r=0$, \autoref{eq:R_Euler} gives
$dP/dr = -(\rho c^2/2)\,d\ln(N^2)/dr = -(\rho c^2)\,N'/N$, i.e. the test-fluid TOV balance in Schwarzschild. Furthermore, for weak fields and small speeds ($N\to 1$, $u_r\simeq v^r$), the gravitational term becomes
$-(c^2/2)\,d\ln(N^2)/dr \simeq -GM/r^2\equiv g$, and \autoref{eq:R_Euler} reduces to the Newtonian form \autoref{eq:NR_Euler}.


\subsection{Radiative diffusion equation} \label{app:sec:relativistic derivation}

In the Eckart frame (i.e., heat flux is orthogonal to the four velocity, $ q_\mu u^\mu = 0 $),  the total energy-momentum tensor is given by
\begin{equation}\label{eq:energy-tensor}
T_{\mu\nu} = (\varepsilon + P)\,u_\mu u_\nu + P\,g_{\mu\nu} + u_\mu q_\nu + u_\nu q_\mu + \pi_{\mu\nu},
\end{equation}
where $\varepsilon$ is the proper energy density (with $\varepsilon \simeq \rho c^2$ in the rest-mass dominated regime used here), $P$ is the isotropic gas pressure, and $\pi_{\mu\nu}$ is the viscous stress tensor. In the slow settling limit $|u^r|\ll 1$ we may neglect $O(u^r)$ corrections in the diffusion closure and take $u^\mu \simeq (N^{-1},\,0,\,0,\,0)$. For spherical symmetry and stationarity configuration we have $q^\mu = (0,\, q^r(r),\, 0,\, 0)$ and the viscous terms $\pi_{\mu\nu}$ vanish by symmetry.

In the diffusion limit, (i.e., mean free path $\lambda \ll$ gradient scale), the radiation mediated heat flux obeys (\cite{1973grav.book.....M}[Equation 22.16i])
\begin{align}\label{eq:heat-flux}
q^\mu = -\kappa_T\, h^{\mu\nu} \left( \nabla_\nu T + T\, a_\nu \right) ,
\end{align}
where $h^{\mu \nu}=g^{\mu \nu}+u^\mu u^\nu$, and $\kappa_T$ is Rosseland mean thermal conductivity
\begin{align}
\kappa_T = \frac{16\sigma T^3}{3 \kappa \rho},
\end{align}
where $\sigma$ is the Stefan–Boltzmann constant, $\kappa$ is the opacity. And, $a_\nu = u^\alpha \nabla_\alpha u_\nu$, where in the static coordinates reduces to 
\begin{align}\label{eq:a-steady}
a_r = \frac{1}{N} \frac{dN}{dr}, \quad 
a_t = a_\theta = a_\varphi = 0.
\end{align}
Because  $u^\mu$ has only a time component, we have  $h^{rr} = g^{rr} = N^2$.
Substituting into the heat flux expression \autoref{eq:heat-flux} gives

\begin{align}\label{eq:q-total derivative}
q^r&= -\frac{16\sigma T^3}{3 \kappa \rho} \, N^2 \left( \frac{\mathrm{d}T}{\mathrm{d}r} + T \frac{1}{N} \frac{\mathrm{d}N}{\mathrm{d}r} \right) \nonumber\\
&= -\frac{16\sigma T^3}{3 \kappa \rho} \, N \left(\frac{\mathrm{d}}{\mathrm{d}r}(N T)\right) .
\end{align}
Using the measured temperature at infinity  $T_\infty \equiv N\, T$, leads to
\begin{align}\label{eq:final_q_ static}
q^r = -\frac{16\sigma T^3}{3 \kappa \rho} \, N \, \frac{\mathrm{d}T_\infty}{\mathrm{d}r}.
\end{align}
In the other hand, in a stationary spacetime with timelike Killing vector $K^\mu=(\partial/\partial t)^\mu$, define the energy current
\begin{equation}
J^\mu \equiv T^{\mu\nu}K_\nu .
\end{equation}
Because $\nabla_\mu T^{\mu\nu}=0$ and $K^\mu$ is Killing,
$\nabla_\mu J^\mu=0$.

Let $S_r$ be the 2-sphere of areal radius $r$ and consider the timelike worldtube $\mathcal{W}_r \equiv S_r \times [t,t+dt]$ with outward unit normal $n_\mu=(0,N^{-1},0,0)$ and induced 3-metric determinant $\sqrt{-h}=N r^2\sin\theta$. The proper 3-surface element on this world-tube is $d\Sigma_\mu = n_\mu\,N\,dA\,dt$, where $dA=r^2\sin\theta\,d\theta d\phi$ is the proper area element on $S_r$. The luminosity measured at infinity (Killing energy flux per unit Killing time) is therefore
\begin{equation}
L_\infty(r)\;\equiv\;-\frac{d}{dt}\int_{\mathcal{W}_r} J^\mu\,d\Sigma_\mu
\,=\,-\,\int_{S_r} J^\mu n_\mu\,N\,dA,
\end{equation}
and Gauss' theorem (using $\nabla_\mu J^\mu=0$) implies that $L_\infty(r)$ is independent of $r$ outside source regions. Strictly, the residual settling flow also carries an advective Killing energy flux. Here we restrict attention to the radiatively dominated regime in which the radial variation of this contribution is subleading relative to $\eta\dot M c^2$. At the working order, we therefore retain the diffusive part of the energy current and treat its corresponding luminosity as radially constant.

Using our stress-energy decomposition \autoref{eq:energy-tensor} with $u^\mu\simeq(N^{-1},0,0,0)$ and purely radial heat flux $q^\mu=(0,q^r,0,0)$ satisfying $u_\mu q^\mu=0$, 
we neglect the energy flux terms proportional to the residual $u\cdot n$ at the working order; with $K\cdot n=0$ and $q\cdot K=0$, the retained diffusive contribution is

\begin{equation}
J^\mu n_\mu=(u\,\cdot\,K)\,(q\,\cdot\,n)=(-N)\,(q^r N^{-1})=-\,q^r.
\end{equation}
Therefore
\begin{equation}\label{eq:lum-ob}
L_\infty(r)=\int_{S_r} N\,q^r\,dA=4\pi r^2\,N\,q^r .
\end{equation}
If $F_{\hat r}$ denotes the local (orthonormal) diffusive flux measured by local static (orthonormal) observers, then $F_{\hat r}=q^{\hat r}=q^r/N$, so $L_\infty=N^2 L_{\rm loc}$ with $L_{\rm loc}=4\pi r^2F_{\hat r}$.

Substituting \autoref{eq:final_q_ static} into \autoref{eq:lum-ob} gives
\begin{align}
L_\infty
&=-\,\frac{64\pi \sigma T^3}{3\,\kappa\,\rho}\, r^2 N^2\,\frac{dT_\infty}{dr}.
\end{align}
Equivalently, eliminating $q^r$ between \autoref{eq:lum-ob} and \autoref{eq:final_q_ static} and solving for the gradient yields
\begin{align}\label{eq:GR-Diff}
\frac{dT_\infty}{dr}
= -\,\frac{3\,\kappa\,\rho}{64\,\pi\,\sigma}\,
\frac{N}{r^{2}\,T_\infty^{3}}\,L_\infty .
\end{align}
Alternative classical derivation also matches also this result (see \autoref{appsec: classic_rad}). Moreover, in a static gravitational field thermal equilibrium implies the Tolman-Ehrenfest relation \citep{tolman_ehrenfest1930,tolman1934}. Furthermore,  in the weak limit $N\to 1$ (so that $T_\infty\to T$ and $L_\infty$ reduces to the standard radiative diffusion law, which is the Newtonian limit of the relativistic heat-flux relation (\cite{1973grav.book.....M}[Equation 22.16i]).

\subsection{Rationalized Form}

It is convenient to absorb redshift factors into the dependent variables and non-dimentionalize with respect to base values $N_f\equiv N(r_f)$, $\rho_f$, and $T_{\rm eq}$:

\begin{subequations}\label{eq:system_bar}
\begin{align}
\bar{x} &= \frac{N^2 \rho}{N_f^2 \rho_f},\\
\bar{y} &= \frac{N T}{T_{\rm eq}} \,=\, \frac{T_\infty}{T_{\rm eq}},\\
s  &= \frac{r-2r_g}{h},
\end{align}
\end{subequations}
with $r_g\equiv GM/c^2$ and the scale 
\begin{equation}
h \,\equiv\, \frac{y_{\rm eq}}{c^2}\,\Big/ \left.\frac{dN}{dr}\right|_{f}
\,\approx\, \frac{kT_{\rm eq}}{m\,|g|},\qquad |g|=\frac{GM}{N_f r_f^2},
\end{equation}
where 
\begin{equation}\label{eq:y-equi}
y_{\rm eq}\equiv k\, T_{\rm eq}/m.  
\end{equation}
For Schwarzschild, $dN/dr=GM/(r^2 c^2 N)$; with $r\simeq r_f$ this gives $h\,\partial_r N \simeq (y_{\rm eq}/c^2)\,(N_f/N)$, hence this matches the NR form up to the extra redshift factor $N_f$. We use the conserved mass flux defined in \autoref{eq:mass_flux},
so that
\begin{equation}\label{eq:co-Ve}
u_r=\frac{u^r}{N^2}=-\,\frac{J}{r^2\,\rho_f\,N_f^2\,\bar{x}}.
\end{equation}
The (Killing energy) luminosity is conserved through the layers, i.e., $L_{\infty}=$ const. Using Equations \eqref{eq:R_continuity}, \eqref{eq:R_Euler}, and \eqref{eq:GR-Diff} with $P=\rho\,kT/m$, and using $\partial_s\equiv h\,\partial_r$ together with Equations \eqref{eq:system_bar}-\eqref{eq:co-Ve}, after replacing $u^r$ and normalizing by $y_{\rm eq}$ one obtains 

\begin{align}\label{eq:R_rEAS}
\partial_s \bar{x}&=\bar{x}\,
\frac{\displaystyle
\underbrace{-\,\frac{N_f}{N}}_{\text{hydrostatic}}
    +
\underbrace{\frac{3\,\bar{y}\,y_{\rm eq}}{c^2}\,\frac{N_f}{N^2}}_{\text{enthalpy correction}}
    +
\underbrace{\frac{3\, \tau_h\,N_f^2}{16N_s^2}\,\frac{L_\infty}{L_{\rm eq}}\,\frac{1}{N}\,
\frac{\bar{x}}{\bar{y}^3}}_{\text{diffusion}}
}
{\displaystyle \bar{y}
 \,-\, \left(\frac{N}{N_f}\right)^3\,\mathcal{T}^2\,
\frac{\bar{y}_f}{\bar{x}^2}} \, .
\end{align}
where $\tau_h$ is given by \autoref{eq:tau_h}, and
\begin{equation}
L_{\rm eq}\equiv \eta\,\dot{M}c^2 \equiv \dfrac{4\pi r_s^2}{N_s^2}\,\sigma\,T_{\rm eq}^4,\qquad N_s\equiv N(r_s). 
\end{equation}
Here, the $1/N$ in the diffusion term is the redshift correction in the relativistic diffusion law, and the factor $N_f^2$ arises because the density variable carried by $\bar{x}$ is the $N^2 \, \rho$ at the base. In a static Schwarzschild atmosphere the proper radial
distance is $dl = dr/N$, so the true optical depth across a thickness
$\Delta r \sim h$ is
\begin{equation}
\tau_{\rm true,h}
= \int_{r_f}^{r_f+h}\kappa\rho\,dl
\simeq \kappa\rho_f \int_{r_f}^{r_f+h}\frac{dr}{N}
\simeq \frac{\kappa\rho_f h}{N_f},
\end{equation}
where we used the thin layer approximation
($\rho\simeq\rho_f$, $N\simeq N_f$). Thus $\tau_h$ differs from
$\tau_{\rm true,h}$ only by the known factor $N_f$; it is a convenient
way of carrying the base optical depth through the algebra.

At the foil ($r = r_f$) we impose the kinetic boundary condition
\begin{equation}\label{eq:base_flux_condition}
    J= \frac{\dot{M}}{4 \pi} \,=\, N_f \, r_f^2 \, \mathcal{T} \, \rho_f \, \sqrt{\frac{kT_f}{m}},
\end{equation}
with $\rho_f\equiv \rho(r_f)$ and $T_f\equiv T(r_f)$.
This is the only place where $\mathcal{T}$ enters; it is the kinetic boundary (mass flux per unit proper area times proper thermal speed) and enforces $|\mathcal{T}| \,< \,1$. Since, using barred variables
\begin{equation}\label{eq:variables-ph-bar}
\rho(r)=\frac{N_f^{2}\rho_f}{N(r)^{2}}\,\bar x(r),
\quad
T(r)=\frac{T_{\rm eq}}{N(r)}\,\bar y(r),
\end{equation}
and $y_{\rm eq}$ is given by \autoref{eq:y-equi}, we note
\begin{equation}
\sqrt{\frac{kT_f}{m}}
=\sqrt{\frac{y_{\rm eq}\,\bar y_f}{N_f}},
\qquad
\bar y_f\equiv \bar y(r_f)=\frac{N_f T_f}{T_{\rm eq}}.
\end{equation}
Hence
\begin{equation}
\label{eq:J-bared}
J= r_f^{2}\,\mathcal{T}\,\rho_f\,N_f^{1/2}\,\sqrt{y_{\rm eq}\,\bar y_f}.
\end{equation}
Therefore, the relativistic sonic term, relativistic analogue of \autoref{eq:NR_rho}, that appears in the denominator of $\partial_s \bar{x}$ in \autoref{eq:R_rEAS} can be written; 
\begin{equation}\label{eq:denominator-px}
D(r)\,\,\equiv\,\underbrace{y_{\rm eq}\,\bar y}_{\displaystyle NkT/m}
\,-\,
\underbrace{\frac{J^{2}}{r^{4}}\,\frac{1}{N\,\rho^{2}}}_{\displaystyle \text{inertial term}}.
\end{equation}
Eliminating $J$ with \autoref{eq:J-bared} and evaluating at the base ($r=r_f$, $N=N_f$, $\bar x_f=1$, $\bar y=\bar y_f$) gives
\begin{equation}
  D_f \equiv {D(r_f)=} \, y_{\rm eq}\,\bar{y}_f\,(1-\mathcal{T}^2).
\end{equation}
Dividing by $y_{\rm eq}$ the common positive factor (as is done when one rationalizes the ODE) gives the dimensionless base discriminant $\bar{y}_f(1-\mathcal{T}^2)$, 
hence the subsonic base condition is precisely
\begin{equation}\label{eq:base-cond-maintext}
    \bar{y}_f(1-\mathcal{T}^2)>0 \quad \Leftrightarrow \quad |\mathcal{T}|<1.
\end{equation}
In addition, the barred diffusion equation has the form

\begin{equation}\label{eq:tem_py}
\partial_s \bar{y}=
-\,\frac{3 \,\tau_h\,N_f^2}{16 N_s^2}\,\frac{L_\infty}{L_{\rm eq}}\,
\frac{1}{N}\,\frac{\bar{x}}{\bar{y}^3},
\end{equation}
so for accretion powered configurations ($L_\infty\le 0$),
$\partial_s\bar y>0$ and $T_\infty$ rises outward, this implies convective stability for the layer.

Further, at the boundary $r=r_f$ fixes $\rho_f$ ( or $J$), and we have
\begin{subequations}\label{eq:R_rBCs}
\begin{align}
&\bar{x}_f = 1,\\
&\bar{y}_f = \frac{N_f T_f}{T_{\rm eq}}.
\end{align}
\end{subequations}
At the top of the layer ($r = r_s$), the luminosity balance with the accretion flow 
\begin{equation}
L_\infty = \frac{4 \pi r_s^2}{N_s^2} \, \sigma T^4_{\infty,s} \, -\, L_{\rm eq}\,=\,L_{\rm eq}\,\big(\bar{y}_s^{4}-1\big)  ,  
\end{equation}
where $T_{\infty,s}=N_s T_s$. Normal stress balance across the interface gives the ram pressure match. On the atmospheric side the normal stress is just the gas pressure,
\begin{equation}
P_s \,=\, \rho_s\,\frac{kT_s}{m}.
\end{equation}
Using the barred definitions \autoref{eq:variables-ph-bar} one finds
\begin{equation}
\label{eq:Ps-bar}
P_s \,=\, \frac{N_f^{2}\rho_f\,y_{\rm eq}}{N_s^{3}}\,\bar x_s\,\bar y_s \, .
\end{equation}
On the accretion side, for a cold, hypersonic upstream the normal momentum flux is
\begin{equation}
T^{r}{}_{r}\,\simeq\,\frac{\rho_a\,(u^r_a)^{2}}{N_s^{2}}.
\end{equation}
Using spherical steady continuity at the shock,
\begin{equation}
\dot M \,=\, -\,4\pi r_s^{2}\,\rho_a\,u^r_a
\quad\Rightarrow\quad
\rho_a=\frac{\dot M}{4\pi r_s^{2}\,|u^r_a|},
\end{equation}
gives
\begin{equation}
\label{eq:Tr_acc}
T^{r}{}_{r}\,\simeq\,\frac{\dot M\,u^r_a}{4\pi r_s^{2}\,N_s^{2}}.
\end{equation}
In a purely ballistic free fall one would take $u^r_a$ at its relativistic value and recover the standard matching. In practice, a fraction of the available binding energy can be radiated away or turned into a wind; the same processes reduce the residual impact speed and hence the incident momentum flux. So by parametrizing the residual impact speed by an impact efficiency
$0<\eta_{\rm ram}\le 1$ via $|u^r_a|\,\equiv\,\eta_{\rm ram}\,c$,  \autoref{eq:Tr_acc} becomes
 
\begin{equation}
\label{eq:Pram}
T^{r}{}_{r}\,\simeq\,\eta_{\rm ram}\,\frac{\dot M\,c}{4\pi r_s^{2}\,N_s^{2}},
\end{equation}
Equating the normal stresses across the thin interface,
$P_s\simeq T^{r}{}_{r}$, and substituting \autoref{eq:Ps-bar}-\autoref{eq:Pram} yields the jump condition
\begin{equation}
\label{eq:barBC-preJ}
\bar x_s\,\bar y_s
\,\simeq\,
\eta_{\rm ram}\,
\frac{\dot M\,c}{4\pi r_s^{2}}\,
\frac{N_s}{N_f^{2}\rho_f\,y_{\rm eq}}.
\end{equation}
To eliminate $\rho_f$ we use the kinetic boundary condition at the foil \autoref{eq:base_flux_condition} which implies
\begin{equation}
\rho_f=\frac{\dot M}{4\pi}\,
\frac{1}{N_f r_f^{2}\mathcal{T}}\,
\sqrt{\frac{m}{kT_f}}
=\frac{\dot M}{4\pi}\,
\frac{1}{r_f^{2}\mathcal{T}}\,
\frac{1}{N_f^{1/2}\,\sqrt{y_{\rm eq}\,\bar y_f}}\,.
\end{equation}
Insert this into \autoref{eq:barBC-preJ} and simplify; using
$|g|h=y_{\rm eq}$ (from $|g|=GM/(N_f r_f^{2})$ and
$h=y_{\rm eq}N_f r_f^{2}/(GM)$) one obtains
\begin{equation}
\label{eq:barBC-final}
\bar x_s\,\bar y_s
\,\simeq\,
\eta_{\rm ram}\,
\frac{\mathcal{T}\,c}{\sqrt{|g|\,h}}\,
\left(\frac{r_f}{r_s}\right)^{\,2}\,
\frac{N_s}{N_f^{3/2}}\,\bar y_f^{1/2}.
\end{equation}
In the thin layer limit $r_s\simeq r_f$, this reduces to
$\bar x_s\bar y_s \simeq \eta_{\rm ram}(\mathcal{T}c/\sqrt{|g|h})\, (N_s/N_f^{3/2})\,\bar y_f^{1/2}$.
In our baseline set-up the upstream just above the atmosphere is a cold, relativistic infall with negligible mass loading and quasi-isotropic radiative losses. In the local static frame at $r_s$, the radial momentum flux and (Killing energy) flux differ by a factor $c$ for a cold stream; hence any process that removes a fixed fraction of the available energy reduces the radial momentum flux by the same fraction. Under these conditions one may identify
$\eta_{\rm ram}=\eta$ (the efficiency appearing in
$L_{\rm eq}=\eta\,\dot M c^{2}$). If mass loading or anisotropic momentum loss is significant upstream, keep $\eta_{\rm ram}$ distinct; all subsequent formulae remain valid with the replacement $\eta\to\eta_{\rm ram}$.

In addition, with $L_\infty \leq 0$ and $kT \ll mc^2$ the relativistic enthalpy correction appears in the numerator of $\partial_s\bar x$, is small. At the base its ratio to the hydrostatic term is
\begin{equation}
\frac{\text{enthalpy}}{\text{hydrostatic}}\Big|_{r_f}
=3\,\frac{kT_f}{m c^2},
\end{equation}
so, it becomes comparable to hydrostatic only if $kT_f\,\sim\, mc^2/3$, independent of $N_f$. Expressed in terms of the redshifted temperature, this is $N_f\lesssim 3\,kT_{\infty,f}/mc^2$, but the controlling quantity is $T_f$. For baryons this would require $T_f\,\gtrsim\,3\times 10^{12}\,$K (protons), far above the temperatures of interest, so the enthalpy term is negligible here.

 It is worth mentioning that the surface redshift $N_f$ enters the rationalized structure equations through known factor $N(r)$ and through the conversion between barred variables and proper quantities $\rho = (N_f^2 \rho_f/N^2)\bar x$, and the optically thick regime the emergent power is deriven to $L_{\rm surf} \simeq\,\eta\dot{M}c^2$ independent of $N_f$. By contrast, local quantities at the base (e.g.\ $\rho_f$, $T_f$ and the physical thickness $h$) do depend on $N_f$, in the near-horizon limit $h\propto N_f$. So the layer becomes geometrically thinner as $N_f\to 0$ even while $L_\infty$ remains fixed.


\subsection{Approximate Solution}
In the limit $r_f\gg r_g$, one has $N\simeq1$ for all $r\ge r_f$ and, for $y_{\rm eq}\ll c^2$, \autoref{eq:R_rEAS} reduces to the nonrelativistic system \autoref{eq:NR_rEAS}; the solutions are then similar. We therefore confine attention to the compact surface limit, $s\,h\ll r_g$, and assume $|L_\infty|$ is small so that a scale separation between $\bar{x}$ and $\bar{y}$ exists (i.e.\ $\bar{y}$ varies weakly across the layer).

\noindent
For $s\equiv(r-2r_g)/h\ll r_g/h$, we obtain
\begin{equation}
 N(r)\,\simeq\,\sqrt{\frac{s\,h}{2r_g}},
\qquad
\frac{N}{N_f}\,\simeq\,\sqrt{\frac{s}{s_f}}\,,
\end{equation}
with $s_f\equiv (r_f-2r_g)/h$ and $N_f=N(r_f)$. Using
$h\equiv (y_{\rm eq}/c^2)/(\partial_r N)_f$ and $(\partial_rN)_f=r_g/(N_f r_f^2)$, one finds
\begin{align}
N_f^2 &\,\approx\, \frac{s_f\,h}{2r_g}
\,=\, \frac{s_f}{2r_g}\,\frac{y_{\rm eq}}{c^2}\,\frac{N_f r_f^2}{r_g}\nonumber\\
&\,\Rightarrow\,
N_f \,\approx\, s_f\,\frac{y_{\rm eq}}{c^2}\,\frac{r_f^2}{2r_g^2}.
\end{align}
In the near-horizon limit $r_f\simeq 2r_g$ this reduces to
\begin{equation}\label{eq:Nf-NH}
    N_f \,\approx\, 2\,s_f\,\frac{kT_{\rm eq}}{m c^2},
\end{equation}
And, correspondingly,
\begin{equation}
h \,=\, \frac{y_{\rm eq}}{c^2}\,\frac{N_f r_f^2}{r_g}
\,\approx\, 4\,N_f\,\frac{kT_{\rm eq}}{m c^2}\,r_g
\,=\, 8\,s_f\,\left(\frac{kT_{\rm eq}}{m c^2}\right)^{\,2}\, r_g,
\end{equation}
so the scale height decreases linearly with $s_f$ (more compact surfaces $\Rightarrow$ smaller $h$).

We consider two regimes:

\medskip
\noindent\textit{Regime A: enthalpy correction negligible}
\ ($kT_f\ll mc^2$, so the positive pressure/enthalpy term in \autoref{eq:R_rEAS} can be dropped).
With $|L_\infty|$ small, take $\bar{y}\simeq\tilde{\bar{y}}$ constant across the layer as a characteristic rationalized temperature and neglect the diffusion term in the $\partial_s\bar{x}$ equation. Then the hydrostatic piece of \autoref{eq:R_rEAS} gives

\begin{equation}
\partial_s \bar{x}\,\simeq\,-\,\frac{\bar{x}}{\tilde{\bar{y}}}\,
\frac{N_f}{N}
=\,-\,\frac{\bar{x}}{\tilde{\bar{y}}}\,
\sqrt{\frac{s_f}{s}}\,.
\end{equation}
Integration yields
\begin{align}
\bar{x}(s) &\,\simeq\,
\exp\,\left[-\,\frac{2}{\tilde{\bar{y}}}\,
\big(\sqrt{s_f s}-s_f\big)\right]\\
&= \rho_f\,\frac{N_f^2}{N^2}\,\bar{x}
= \rho_f\,\frac{s_f}{s}\,
\exp\,\left[-\,\frac{2\,s_f}{\tilde{\bar{y}}}\,
\left(\sqrt{\frac{s}{s_f}}-1\right)\right]\nonumber.
\end{align}
Thus, the density (sub)exponentially decreases. As comparing with NR case \autoref{xbarNR} in the NR atmosphere (small redshift, $r_f\gg r_g$ gravity is essentially constant over a scale height \autoref{eq:Hassump}. We have defined 
\begin{equation}
 s_{\rm NR}=\frac{r-r_f}{H}.   
\end{equation}
The hydrostatic equation in that limit becomes (with almost constant $T$) 
\begin{equation}
    \frac{d \rho}{dr} \simeq -\frac{\rho}{H}, \quad \Rightarrow \quad  \frac{d \ln \rho}{ds_{\rm NR}} \simeq -1,
\end{equation}
so 
\begin{equation}
 \bar{x}_{\rm SN}(s_{\rm NR})=\frac{\rho}{\rho_f} \simeq \,\exp\,[-s_{\rm NR}].  
\end{equation}
In GR part we defined $s=(r-2r_g)/h$ and then used the near-horizon relation $N\sim \sqrt{s}$ to pull all redshift dependence into $s$, so the same symbol $s$ is not the same "linear height" as in the NR section; it is matched to the GR geometry. Thus, in NR, the local hydrostatic balance sees a constant gravitational acceleration, so $\rho$ decays like an ordinary exponential with height. In the near-horizon GR limit, the gravitational redshift factor $N$ changes as $\sqrt{r-2r_g}$ and we see that directly in the $\sqrt{s}$ that appears in the exponent.

Expanding
$\sqrt{s/s_f}=1+(s-s_f)/(2s_f)$, the local e-folding thickness in $s$ is
\begin{equation}
 \Delta s_{\rm efold}
=\left(\frac{1}{s_f}+\frac{1}{\tilde{\bar y}}\right)^{-1}
\simeq \tilde{\bar y},\qquad (s_f\gg \tilde{\bar y}),   
\end{equation}
near the base, so the physical thickness is 
\begin{equation}
\Delta r\simeq h\,\Delta s_{\rm efold}\propto s_f.   
\end{equation}
Hence, more compact surfaces (smaller $s_f$) have a thinner atmosphere, consistent with what we had earlier that $h$ decreasing linearly with $s_f$.

The associated temperature profile follows from the relativistic diffusion equation in \autoref{eq:R_rEAS}.
With $\bar{x}\simeq\tilde{\bar{x}}$ and $N^{-1}\simeq\sqrt{2r_g/(s\,h)}$,
\begin{align}
\partial_s \bar{y}\,\simeq& -\,\frac{3\,\tau_h\,N_f^2}{16N_s^2}\,\frac{L_\infty}{L_{\rm eq}}\,
\frac{1}{N}\,\frac{\tilde{\bar{x}}}{\bar{y}^3}\nonumber\\
&=\,-\,\frac{3\,\tau_h\,N_f^2}{16N_s^2}\,\frac{L_\infty}{L_{\rm eq}}\,
\sqrt{\frac{2r_g}{s\,h}}\,\frac{\tilde{\bar{x}}}{\bar{y}^3}.
 \end{align}
Thus
\begin{equation}\label{eq:R_yapprox}
\bar{y}(s)= \left[
\bar{y}_f^4- \frac{3}{2}\,\bar{\tau}\,\frac{N_f^2}{N_s^2}\,\frac{L_\infty}{L_{\rm eq}}\,
\sqrt{\frac{2r_g}{h}}\,\big(\sqrt{s}-\sqrt{s_f}\big)\right]^{1/4},
\end{equation}
where $\bar{\tau}\equiv\tau_h\,\tilde{\bar{x}}$ and $\tau_h$ is given by \autoref{eq:tau_h}.

\medskip

\textit{Surface luminosity.} Using $L_\infty=L_{\rm eq}(\bar{y}_s^4-1)$, and the integrated profile for $\bar y^4$, we obtain 
\begin{align}
L_{\rm surf}\,&\equiv\,L_{\rm eq}\,\bar{y}_s^4\\
&= \frac{\,L_{\rm eq}\,\bar{y}_f^4
 + \frac{3}{2}\,\bar{\tau}\,\frac{N_f^2}{N_s^2}\,L_{\rm eq}\,\sqrt{\frac{2r_g}{h}}\,
(\sqrt{s_s}-\sqrt{s_f})\,}{\,1 + \frac{3}{2}\,\bar{\tau}\,\frac{N_f^2}{N_s^2}\,\sqrt{\frac{2r_g}{h}}\,
(\sqrt{s_s}-\sqrt{s_f})\,}\nonumber\,.
\end{align}
Since $\bar y_f=T_{\infty,f}/T_{\rm eq}=N_f T_f/T_{\rm eq}$ and $r_s\simeq r_f$, the base term is
\begin{align}
L_{\rm eq}\,\bar y_f^4 &\,=\, \frac{4\pi r_f^2}{N_f^2}\,\sigma\,T_{\infty,f}^4\\
&\,=\, 4\pi r_f^2\,\sigma\,N_f^2\,T_f^4
\,\equiv\, L_{\rm foil,\infty}\nonumber.
\end{align}
Equivalently, if one prefers the local definition $L_{\rm foil}=4\pi r_f^2\sigma T_f^4$, then
$L_{\rm eq}\bar y_f^4=N_f^2\,L_{\rm foil}$. In either form,
\begin{equation}
L_{\rm surf}
= \frac{\,L_{\rm foil,\infty}
 + \frac{3}{2}\,\bar{\tau}\,\frac{N_f^2}{N_s^2}\,L_{\rm eq}\,\sqrt{\frac{2r_g}{h}}\,
(\sqrt{s_s}-\sqrt{s_f})\,}{\,1 + \frac{3}{2}\,\bar{\tau}\,\frac{N_f^2}{N_s^2}\,\sqrt{\frac{2r_g}{h}}\,
(\sqrt{s_s}-\sqrt{s_f})\,}\,.
\end{equation}
Thus, even for very compact surfaces one obtains $L_{\rm surf}\to L_{\rm eq}=\eta\dot{M}c^2$ i.e., it radiates at the equilibrium rate.
To make it clear let us compare it to NR case. Here the role of $(3/4)\bar\tau$ in \autoref{eq:Lsurf-approx} is played by the whole combination 

\begin{equation}
 A \,\equiv \frac{3}{2}\bar{\tau}\,\frac{N_f^2}{N_s^2}\,\sqrt{\frac{2r_g}{h}}\,
(\sqrt{s}-\sqrt{s_f}),  
\end{equation}
Then our GR expression is just
\begin{equation}
L_{\rm surf}
\,=\,\frac{\,L_{\rm foil} + \displaystyle A L_{\rm eq}}{1+\displaystyle A}.
\end{equation}
So to keep this similarity, we define dimensionless combination

\begin{equation}
 \bar{\tau}_{\rm GR} \,\equiv 2\,\bar{\tau}\,\frac{N_f^2}N_s^2\,\sqrt{\frac{2r_g}{h}}\,
( \sqrt{s}-\sqrt{s_f} ),  
\end{equation}
the $2$ is the only new geometric factor from integrating $1/N \propto s^{-1/2}$ in the near-horizon geometry, and the extra $4/3$ is just a choice of normalization so that the GR and NR $L_{\rm surf}$ formulas \autoref{eq:Lsurf-approx} look identical. Thus, we obtain
\begin{equation}
L_{\mathrm{surf}}
= \frac{L_{\mathrm{foil},\infty}+\frac{3}{4}\,\bar\tau_{\mathrm{GR}}\, L_{\mathrm{eq}}}
{1 + \frac{3}{4}\,\bar \tau_{\mathrm{GR}}}\,,
\end{equation}
and the optically thick condition is just
\begin{align}
\frac{3}{4} \bar{\tau_{\rm GR}} \gg 1.
\end{align}
It is worth mentioning that, despite its algebraic complexity, $\bar{\tau}_{\rm GR}$ is the redshift-weighted optical depth combination that controls radiative diffusion in the rescaled GR variables. Thus, $\bar{\tau}_{\rm GR}\gg1$ is the appropriate diffusion-dominated condition for the settling layer considered here.


\noindent\textit{Regime B: enthalpy correction important}
\ (rare in practice). The positive pressure/enthalpy term in the numerator of \autoref{eq:R_rEAS} competes with the hydrostatic term when
$3\,\frac{kT_f}{m c^2}\sim 1$.
This criterion is independent of $N_f$; it requires ultra-relativistic baryon temperatures and is not met for the sources of interest.

Formally keeping this term and still neglecting diffusion (small $|L_\infty|$), and taking the subsonic-base limit so the denominator is $\simeq \bar y\simeq \tilde{\bar y}$, we obtain
\begin{equation}
\partial_s \bar{x}\simeq \bar{x}\left[\frac{3\,y_{\rm eq}}{c^2}\,\frac{N_f}{N^2}
 - \frac{1}{\tilde{\bar{y}}}\frac{N_f}{N}\right].
\end{equation}
Using $N_f/N=\sqrt{s_f/s}$ and, in the near-horizon limit,
\begin{equation}
\frac{N_f}{N^2}=\frac{2r_g N_f}{h}\,\frac{1}{s}
=\frac{c^2}{2y_{\rm eq}}\,\frac{1}{s},
\end{equation}
this reduces to
\begin{equation}
\partial_s\ln\bar{x}\,\simeq\,\frac{3}{2}\,\frac{1}{s}\,-\,\frac{1}{\tilde{\bar y}}\sqrt{\frac{s_f}{s}}.
\end{equation}
Integrating gives
\begin{equation}
\bar{x}(s)\,\simeq\,
\left(\frac{s}{s_f}\right)^{3/2}\,
\exp\,\left[-\,\frac{2\,\sqrt{s_f}}{\tilde{\bar{y}}}\,
\big(\sqrt{s}-\sqrt{s_f}\big)\right].
\end{equation}
When the enthalpy term dominates near the base (i.e.\ $3kT_f/mc^2>1$), $\bar{x}$ (and thus $\rho$) initially increases with radius; the pressure nevertheless decreases outward because $P\propto \bar{x}/N^3$ and $N\propto \sqrt{s}$ near the horizon, so the $s^{3/2}$ rise is canceled by the geometric $N^{-3}$ factor, leaving the hydrostatic exponential to enforce $dP/dr<0$.
The density peaks where $d\ln\bar{x}/ds=0$, which gives
\begin{equation}\label{eq:speak:)}
    s_{\rm peak} \,=\, \frac{9}{4}\,\frac{\tilde{\bar{y}}^{2}}{s_f}.
\end{equation}
Note that since near the horizon using \autoref{eq:Nf-NH}, one has $s_f\propto N_f=(1+z_f)^{-1}$, and therefore
$s_{\rm peak}\propto s_f^{-1}\propto N_f^{-1}\propto (1+z_f)$. Hence the density peak is pushed outward as the surface redshift increases,
which is a qualitatively different behavior from the Newtonian case.

From \autoref{eq:speak:)} we see, at the peak (and in the compact limit $s_{\rm peak}\gg s_f$) we obtain
\begin{equation}
    \bar{x}_{\rm peak}
    \,\simeq\,
    \left(\frac{s_{\rm peak}}{s_f}\right)^{3/2}
    \exp (-3)
    =\frac{27}{8}\frac{\tilde{\bar{y}}^{3}}{s_f^3}e^{-3}.
\end{equation}
Setting $\tilde{\bar{x}}=\bar{x}_{\rm peak}$, the temperature is approximately given by \autoref{eq:R_yapprox} (with the corrected coefficient in that equation), with the attendant consequences for $L_{\rm surf}$.

Physically, in Regime B the gas is relativistically hot, i.e., the relativistic enthalpy and inertia terms are comparable those associated with gravity. Nevertheless, the local temperature $T=T_{\infty}/N$ drops outward as $N$ rises, and does so sufficiently fast that $P=\rho kT/m$ decreases even if $\rho$ increases with radius initially before ultimately falling.

\section{Boundary Condition Considerations}\label{sec:boundary}

There are a number of constraints on the parameters imposed by the boundary conditions. We review these here in terms of convenient parametrizations of the foil surface properties.  We note that we assume locality, i.e., that the foil surface can only interact with matter that impinges upon it.

\subsection{Mass Flow}
Already, we have explicitly noted the need for
\begin{equation}
\mathcal{T}^2<1,
\end{equation}
limiting the rate at which the foil can absorb accreting baryonic matter. (Black holes explicitly violate this condition.)  This implicitly imposes conditions on the surface redshift.

Because the pressure is monotonically decreasing with radius in our solutions, the pressure at the base must be at least the post shock pressure:
\begin{equation}
\begin{gathered}
P_f\,\,=\, \frac{\dot{M}\,y_{\rm eq}^{1/2}}{4\pi r_f^2 \mathcal{T}}\,
\frac{\bar{y}_f^{1/2}}{N_f^{3/2}}
\,\ge\, \frac{\eta\,\dot{M}\,c}{4\pi r_s^2 N_s^2}\,,
\end{gathered}
\end{equation}
This gives a lower bound on $\bar{y}_f$ and hence on $T_f (=\bar{y}_f T_{\rm eq}/N_f)$:
\begin{align}
\bar{y}_f \,&\ge\, \eta_{\rm ram}^{2}\,\mathcal{T}^{2}
\left(\frac{r_f}{r_s}\right)^{4}
\frac{N_f^{3}}{N_s^{4}}\,
\frac{m c^{2}}{kT_{\rm eq}}\\
&\quad\Longleftrightarrow\quad
T_f \,\ge\, \eta_{\rm ram}^{2}\,\mathcal{T}^{2}
\left(\frac{r_f}{r_s}\right)^{4}
\frac{N_f^{2}}{N_s^{4}}\,\nonumber
\frac{m c^{2}}{k}.
\end{align}
As $\mathcal{T}\to 0$, this bound allows $T_f\to 0$.
Additionally, the atmosphere must be optically thick over a local scale height (as assumed by the diffusion closure), i.e.\ $\tau>1$.
For electron scattering,
\begin{equation}
\tau \,\approx\, \kappa\,\rho\,\frac{kT}{m|g|}
\,=\, \frac{\kappa\,P}{|g|}\,,
\end{equation}
so (since $P$ decreases outward)
$\tau \lesssim \kappa P_f/|g|$, and we must have
\begin{equation}
P_f \,=\, \frac{\dot{M}\,y_{\rm eq}^{1/2}}{4\pi r_f^2 \mathcal{T}}\,
\frac{\bar{y}_f^{1/2}}{N_f^{3/2}}\,>\, \frac{|g|}{\kappa}\,.
\end{equation}
Solving for $T_f$ gives
\begin{equation}\label{T_f-bound}
T_f \,>\, \frac{m}{k}\,\mathcal{T}^2
\left(\frac{4\pi\,|g|\,r_f^2}{\kappa\,\dot{M}}\right)^{\,2}\,N_f^{2}\,.
\end{equation}
Using the height scale definition \autoref{eq:h-equib} (which implies $|g|r_f^2 = GM/N_f$), the explicit $N_f$ cancels and we obtain
\begin{equation}
T_f \,>\, \frac{m}{k}\,\mathcal{T}^2
\left(\frac{4\pi GM}{\kappa\,\dot{M}}\right)^{\,2}.
\end{equation}
Equivalently, in terms of $L_{\rm eq}$ and the Eddington luminosity
$L_{\rm Edd}=4\pi GM c/\kappa$,
\begin{equation}
T_f \,>\, \frac{m c^2}{k}\,\mathcal{T}^2\,\eta^2
\left(\frac{L_{\rm Edd}}{L_{\rm eq}}\right)^{\,2},
\end{equation}
where we used $L_{\rm eq}=\eta\,\dot{M}\,c^2$.

Which bound on $T_f$ is more restrictive depends on $\dot{M}$ (or $L_{\rm eq}$). The shock support bound scales as $T_f\propto N_f^2$, whereas the optical depth bound is independent of $N_f$. Thus the shock bound dominates only when
\begin{equation}
\left(\frac{r_f}{r_s}\right)^2\frac{N_f}{N_s^2}
\,>\,\frac{L_{\rm Edd}}{L_{\rm eq}}\,,
\end{equation}
which, for typical Radiatively inefficient accretion flows (RIAFs) conditions $L_{\rm eq}\ll L_{\rm Edd}$, would require extremely large surface redshifts (often unattainable).

Finally, the emitting layer redshift is limited by the requirement that the solution extends beyond the density peak $s_{\rm peak}$.
Using the near horizon relation $N^2\simeq s h/(2r_g)$ and the Regime B peak \autoref{eq:speak:)},
\begin{align}
    N_s^2 \,=\, \frac{h \,s_s }{2r_g}
&\,>\, \frac{h\,s_{\rm peak}}{2r_g}\\
&\,=\, 4\,s_{\rm peak}\,s_f\left(\frac{kT_{\rm eq}}{m c^2}\right)^{\,2}
\,=\, 9\,\tilde{\bar y}^{\,2}\left(\frac{kT_{\rm eq}}{m c^2}\right)^{\,2}.\nonumber
\end{align}
Hence
\begin{equation}
N_s \,\gtrsim\, 3\,\tilde{\bar y}\,\frac{kT_{\rm eq}}{m c^2}.
\end{equation}
For $kT_{\rm eq}/(mc^2)\ll 1$ (e.g.\ $T_{\rm eq}\sim 10^4$ K for baryons), this lower bound on $N_s$ is tiny, i.e. it is a very weak constraint on the photosphere redshift.


\subsection{Energy Flow}
\label{sec:microBC}

Up to now we treated the gas temperature at the base of the atmosphere, $T_f\equiv T(r_f)$, as a free inner datum. 
However, local gas-surface microphysics constrains both $T_f$ and the base temperature gradient $\partial_rT|_{f}$:  even if the foil itself is extremely cold, baryon-baryon collisions and radiative diffusion generally prevent the base gas from cooling arbitrarily, producing an effective floor temperature. We model two local ingredients.

\paragraph{(i) Energy accommodation in gas--surface collisions.}
Let $T_\bullet$ denote the foil material temperature. 
A standard way to parametrize partial thermalization at the wall is with an energy accommodation coefficient $\epsilon\in[0,1]$ defined by
\begin{equation}\label{eq:epsilon}
T_{\rm refl} \,=\, T_f \,+\,\epsilon\,(T_\bullet-T_f),
\qquad
\epsilon\in[0,1],
\end{equation}
where $T_{\rm refl}$ is the effective temperature of particles immediately after reflection/interaction with the surface. 
For a simple elastic 1D collision model between a gas particle of mass $m$ and a surface site of effective mass $M$ (thermalized at $T_\bullet$), one finds
\begin{equation}\label{eq:epstin}
\epsilon=\frac{4mM}{(m+M)^2},
\end{equation}
so $\epsilon\to 0$ corresponds to nearly specular reflection ($M\to\infty$) and $\epsilon=1$ corresponds to complete thermal accommodation ($m=M$).

\paragraph{(ii) Heat-flux balance at the wall.}
The net energy flux transferred from the foil to the gas (positive when energy flows from foil to gas) can be estimated kinetically as the particle flux incident on the wall times the energy gained per particle:
\begin{align}
q_{\rm wall}\, \simeq& \frac{1}{4}\,n\,c_v\,v_{\rm th}\,(T_{\rm refl}-T_f)\nonumber\\
&= \frac{1}{4}\,n\,c_v\,\epsilon\,v_{\rm th}\,(T_\bullet-T_f),
\end{align}
where $c_v=\frac{3}{2}k$, $n$ is the baryon number density and $v_{\rm th}$ is a characteristic thermal speed.
Within the gas, the collisional conductive flux is
\begin{equation}
q \,=\, -\,\kappa_c\,\partial_r T,
\qquad
\kappa_c=\frac{1}{3}\,n\,c_v\,v_{\rm th}\,\lambda_c,
\end{equation}
with $\lambda_c$ the baryon-baryon collisional mean free path near the wall.
Matching $q_{\rm wall}=q$ at $r=r_f$ eliminates $n,c_v,v_{\rm th}$ and yields a kinetic boundary condition on the base temperature gradient:
\begin{equation}
\partial_r T\Big|_f
= -\,\frac{3\,\epsilon}{4\,\lambda_c}\,(T_\bullet-T_f).
\label{eq:BC_kinetic}
\end{equation}
On the macroscopic side, the base temperature gradient is also set by radiative diffusion in the optically thick atmosphere. 
Using the same diffusion law as in the main text (evaluated locally over a microscopic distance $\sim\lambda_c$, across which $N(r)$ is constant to excellent approximation),
\begin{equation}\label{eq:BC_diffusive}
\partial_r T\Big|_f
= -\,\frac{3\,\kappa\rho}{64\pi\sigma}\, 
\frac{L}{r_f^2\,T_f^{3}}
\,=\,-\,\frac{3}{64\pi\sigma}\,
\frac{L}{r_f^2\,\lambda\,T_f^{3}},
\end{equation}
where $\lambda\equiv(\kappa\rho)^{-1}$ is the photon mean free path and $L$ is the (radially conserved) luminosity (outward positive).
Equating \autoref{eq:BC_kinetic} and \autoref{eq:BC_diffusive} gives an implicit local relation for $T_f$:
\begin{equation} \label{eq:Tf_floor}
T_f^{3}\,(T_f-T_\bullet)
\,=\,\frac{T_{\rm eq}^{4}}{4\,\epsilon}\,
\frac{\lambda_c}{\lambda}\,\frac{|L|}{L_{\rm eq}},
\end{equation}
where $L_{\rm eq}\equiv 4\pi r_f^2\sigma T_{\rm eq}^4$.
(The absolute value is convenient because in accretion-powered configurations the net luminosity through the layer is typically inward, $L\le 0$.)

It is convenient to define the microphysical floor temperature
\begin{equation}\label{eq:TL_def}
T_L \,\equiv\,T_{\rm eq}
\left(\frac{1}{4\,\epsilon}\,\frac{\lambda_c}{\lambda}\,\frac{|L|}{L_{\rm eq}}\right)^{1/4},
\end{equation}
so that \autoref{eq:Tf_floor} becomes
\begin{equation}
\label{eq:dimless_cubic}
\left(\frac{T_f}{T_L}\right)^{\,3}\left(\frac{T_f}{T_L} - \frac{T_\bullet}{T_L}\right) \,=\, 1.
\end{equation}
Defining $t_f\equiv T_f/T_L$ and $s_\bullet\equiv T_\bullet/T_L$, \autoref{eq:dimless_cubic} has a unique positive root with limits
\begin{align}
&s_\bullet\ll 1\,:\,
t_f \,=\, 1 + \frac{s_\bullet}{4} + O(s_\bullet^2)
\quad\Rightarrow\quad T_f \simeq T_L,\\
&s_\bullet\gg 1\,:\,
t_f \,=\, s_\bullet + \frac{1}{s_\bullet^3} + O(s_\bullet^{-7})
\quad\Rightarrow\quad
T_f \simeq T_\bullet + \frac{T_L^4}{T_\bullet^3}\nonumber,
\end{align}
so $T_f$ approaches $T_\bullet$ from above for hot foils, while for cold foils ($T_\bullet\to 0$) the base gas temperature cannot drop below $T_L$.

Because $N(r)$ varies negligibly across the microscopic thickness $\Delta r\sim\lambda_c$, the same boundary condition applies in the relativistic treatment when expressed in redshift-invariant variables: replacing $(T,L,T_{\rm eq},L_{\rm eq})$ by $(T_\infty,L_\infty,T_{{\rm eq},\infty},L_{{\rm eq},\infty})$ multiplies both sides of \autoref{eq:Tf_floor} by the same factor and leaves the relation unchanged up to ${\cal O}(\lambda_c/\ell_g)\ll1$, with $\ell_g\equiv N/|dN/dr|$.

\begin{figure}[t]
  \centering
  \includegraphics[width=\columnwidth]{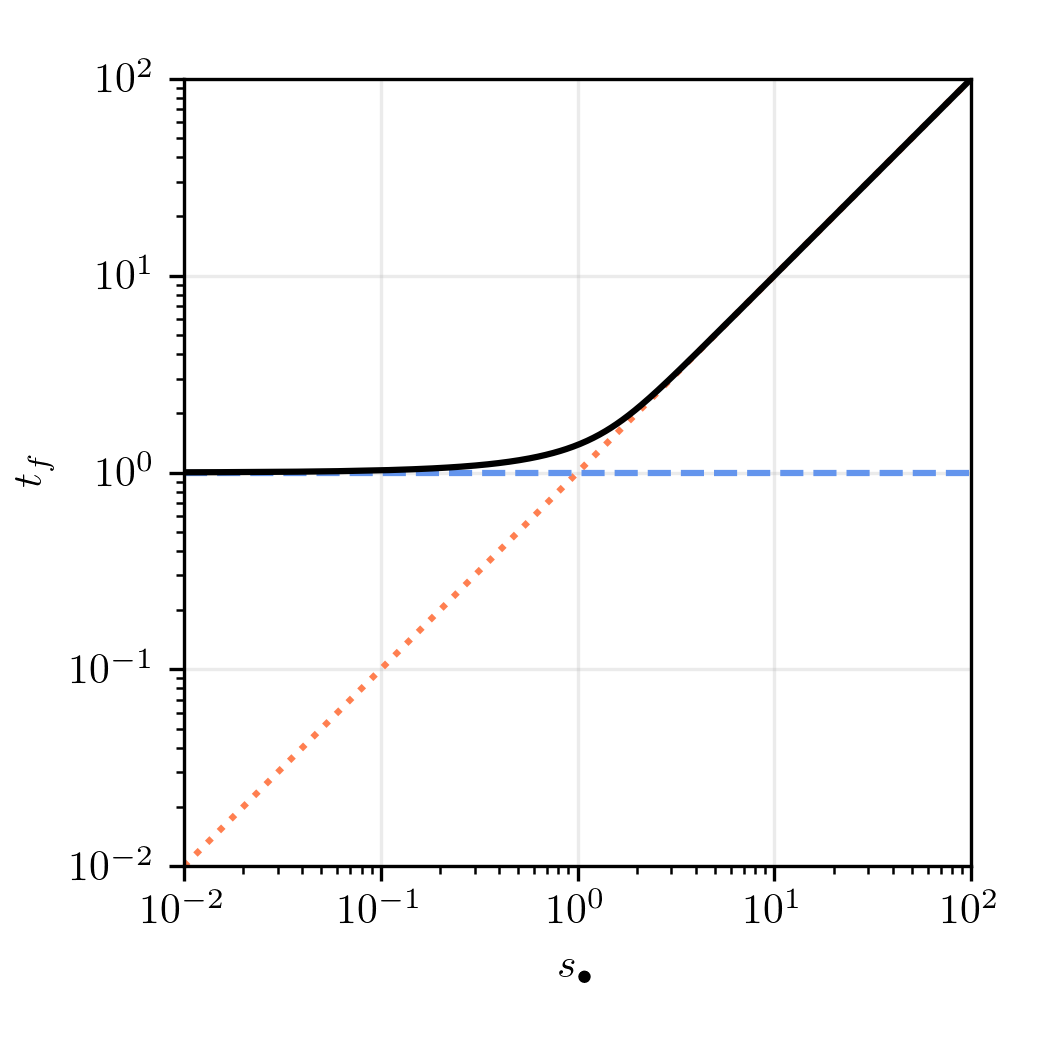}
  \caption{Microphysical floor on the base temperature from \autoref{eq:dimless_cubic} on log-log axes.
  Blue curve shows the solution of $t^{3}(t-s_\bullet)=1$ with $t_f\equiv T_f/T_L$ and
  $s_\bullet\equiv T_\bullet/T_L$. Black dashed line shows $t_f=1$ ($T_f=T_L$). Red dotted line shows $t_f=s_\bullet$ ($T_f=T_\bullet$). For $s_\bullet \gg 1$ (hot foil), $t_f\to s_\bullet$ with a small upward offset $T_L^4/T_\bullet^3$. Thus $T_L$ sets a hard lower bound for the gas temperature at the base; it depends on $\epsilon$, $\lambda_c/\lambda$, and the normalized luminosity $|L|/L_{\mathrm{eq}}$ via \autoref{eq:TL_def}, while $\lambda_c/\lambda$ and $L$ are being held fixed in this specific accommodation model.}
  \label{fig:Tf_floor}
\end{figure}

\subsection{Analytic sensitivity to the energy accommodation coefficient $\epsilon$}
\label{sec:epsilon|sensitivity}

we already saw that  (\autoref{sec:microBC}) the accommodation coefficient $\epsilon\in[0,1]$, \autoref{eq:epstin}, enters only through the microscopic gas-surface heat exchange boundary condition via \autoref{eq:epsilon}. In this subsection, we quantify its effect analytically by holding fixed the outer supply parameters
that set the luminosity crossing the base ($L$ or $L_\infty$),
and fixed the microphysical ratio of $\lambda_c/\lambda$.

\medskip

For the accretion powered configurations, emphasized in this work, the net luminosity in our sign convention satisfies $L\le 0$. And therefore the diffusion gradient at the base is outward ($\partial_r T|_f>0$) and the gas emerges hotter than the foil material, $T_f>T_\bullet$. From \autoref{eq:TL_def} where we have defined the (positive) floor scale by $T_L$, we obtain
\begin{equation}\label{eq:TL_def_abs}
\frac{d\ln T_L}{d\ln\epsilon}=-\frac{1}{4}.
\end{equation}
Now we define
\begin{equation}
F(T_f,\epsilon)\;\equiv\;T_f^4-T_\bullet T_f^3-T_L^4(\epsilon)=0,
\end{equation}
and we know $(T_L^4(\epsilon)\propto \epsilon^{-1})$. Differentiating implicitly at fixed $T_\bullet$, $|L|/L_{\rm eq}$, and $\lambda_c/\lambda$ gives
\begin{equation}
\left(4T_f^3-3T_\bullet T_f^2\right)\,dT_f \;=\; d\,\left(T_L^4\right)
\;=\; -\,T_L^4\,d\ln\epsilon,
\end{equation}
and by using $T_L^4=T_f^3(T_f-T_\bullet)$ from \autoref{eq:Tf_floor} yields
\begin{equation}
\frac{d\ln T_f}{d\ln\epsilon}=-\frac{T_f-T_\bullet}{4T_f-3T_\bullet}.
\label{eq:dlnTf_dlnEps}
\end{equation}
On the relevant physical branch ($T_f\ge T_\bullet$), the denominator is positive
($4T_f-3T_\bullet \ge T_f>0$), and thus 
\begin{equation}\label{eq:dlnTf_bounds}
0 \;\ge\; \frac{d\ln T_f}{d\ln\epsilon} \;\ge\; -\frac14.
\end{equation}
Thus, $T_f$ depends on $\epsilon$ at most as a fourth root. In particular,
\begin{align}
T_f \gg T_\bullet:&\quad \frac{d\ln T_f}{d\ln\epsilon}\to -\frac14
\ \Rightarrow\ T_f\propto \epsilon^{-1/4},\\
T_f\to T_\bullet:&\quad \frac{d\ln T_f}{d\ln\epsilon}\to 0
\ \Rightarrow\ T_f\simeq T_\bullet\ \text{(weak $\epsilon$-dependence)}.
\end{align}
As a concrete example, even a large change in $\epsilon\to 10^{-4}\epsilon$, will change $T_f$ by at most a factor $10$ (and typically less, unless $T_\bullet\ll T_f$).

\subsubsection{Propagation to base density and optical depth}

Through the mass-flux boundary condition, the base density scales as
$\rho_f\propto v_{\rm th,f}^{-1}\propto T_f^{-1/2}$, and thus
\begin{align}\label{eq:dlnRhof_dlnEps}
\frac{d\ln\rho_f}{d\ln\epsilon}&=-\frac12\,\frac{d\ln T_f}{d\ln\epsilon}\nonumber\\
&=\frac{T_f-T_\bullet}{2(4T_f-3T_\bullet)}\;\in\,\left[0,\frac{1}{8}\right].
\end{align}
If one estimates the transport optical depth across a local scale height using
$H\propto T_f$ and $\tau_H\sim \kappa_R\rho_f H\propto \rho_f T_f\propto T_f^{1/2}$, then

\begin{align}\label{eq:dlnTauH_dlnEps}
\frac{d\ln\tau_H}{d\ln\epsilon}
&=\frac12\,\frac{d\ln T_f}{d\ln\epsilon}\nonumber\\
&=-\frac{T_f-T_\bullet}{2(4T_f-3T_\bullet)}\in\left[-\frac18,0\right].
\end{align}
Therefore, within the valid region, $\epsilon$ produces only weak changes in $\rho_f$ and in the diffusion measure $\tau_H$.

Also, in the NR approximate expression \eqref{eq:Lsurf-approx}, for example, one has

\begin{equation}
\frac{\partial L_{\rm surf}}{\partial L_{\rm foil}}=\frac{1}{1+\frac34\bar\tau},
\end{equation}
so any $\epsilon$ dependence entering through $L_{\rm foil}\propto T_f^4$
is weighted by the small factor $(1+\tfrac34\bar\tau)^{-1}$ when $\bar\tau\gg 1$.

\section{Summary and Conclusions}
\label{sec:conclusions}

We have constructed a minimal model of a steady, spherically symmetric, subsonic baryonic settling layer between an outer photosphere (or accretion shock) and a horizonless compact surface (a ``foil'') embedded in an exterior Schwarzschild spacetime. The layer is gas-pressure supported and transports energy by radiative diffusion in the optically thick limit. We use the nonrelativistic equations to build intuition and then formulate the full problem relativistically in terms of redshift-invariant observables.

The unknown surface physics is encoded through local boundary conditions at the foil. In particular, we parametrize the rate at which baryons are processed by the surface with a dimensionless transmittance parameter $\mathcal{T}$ and allow the foil temperature to differ from the gas temperature. Subsonic settling restricts $\mathcal{T}$ to be below unity in magnitude. At the outer interface we impose (i) energy balance with the accretion flow (setting the conserved luminosity at infinity) and (ii) normal-stress balance with the incoming ram pressure. Two additional viability requirements arise within the settling layer: the base pressure must be sufficient to match the post-shock pressure, and the atmosphere must be optically thick across a local scale height so that diffusion is self-consistent. For sub-Eddington supply the optical-thickness requirement typically provides the more restrictive lower bound on the base temperature; making the shock-support bound dominant would require an extreme redshift hierarchy between the foil and the emitting photosphere.

For compact surfaces the solutions exhibit a clean scale separation: the redshifted temperature varies only weakly across the layer while the density decreases rapidly (more slowly than a simple exponential in the natural near-horizon height coordinate). The physical thickness of the atmosphere shrinks as the surface becomes more compact. A relativistic enthalpy correction can modify the immediate base behavior only at ultra-relativistic baryon temperatures (well above those relevant for the applications of interest); in the parameter regime considered here the layer remains pressure supported, convectively stable, and globally subsonic.

Because the relevant Killing energy current is conserved, the luminosity measured at infinity is radius independent throughout the atmosphere. Consequently, in the optically thick diffusion dominated regime the emergent photosphere luminosity is driven close to the accretion power processed through the layer, largely independent of the foil redshift. Within the assumptions adopted here (steady state, spherical symmetry, local surface interactions, and diffusive transport in an optically thick scattering atmosphere), redshift and diffusion do not provide a generic mechanism to hide accretion power from distant observers.

Finally, the microphysics of gas-surface interaction imposes an additional local constraint at the base: gas-surface collisions and conductive transport produce an effective floor on the base gas temperature that depends on the accommodation efficiency and the ratio of microscopic mean free paths, together with the luminosity crossing the base. This replaces the assumption of an arbitrary base temperature with a controlled, local boundary condition.

Effects beyond the present scope (rotation, magnetic stresses, departures from spherical symmetry, and composition-dependent opacities) can be incorporated as extensions. While these can change detailed profiles and the location of the photosphere, they are not expected to alter the central conservation argument that ties the emergent luminosity to the accretion power in optically thick settling layers. The observational consequences of this result can be compared directly with existing limits on the thermal surface emission. For \SgrA millimeter and infrared observations constrain the luminosity of any visible compact surface to less than approximately $0.4\%$ of the inferred accretion luminosity \cite{2009ApJ...701.1357B}. Optical and near infrared limits on \VirA combined with the minimum accretion rate required to power its jet, likewise strongly constrain a thermally emitting surface \cite{2015ApJ...805..179B}. At stellar masses, the systematically lower quiescent luminosities of black hole X-ray novae relative to comparable neutron star systems provide and analogous test \cite{1997ApJ...478L..79N,2001ApJ...553L..47G}. The present calculation supplies the complementary theoretical step as if a foil develops the optically thick baryonic setteling atmosphere presented in this work, then the accretion power emerges as thermal photospheric emission; therefore, current observational upper limits on such emission, constrain the broader class of locally interacting foil models considered in this paper. A source-specific application of this framework will require determining the region of $(\mathcal{T},\epsilon,T_\bullet)$ space for which the atmosphere becomes optically thick, comparing its formation time $t_{\rm form}\sim\max\,(t_{\rm set},t_{\rm diff})$ with the accretion and variability timescales of the source, and then applying existing limits on thermal surface emission. In particular, it is interesting to have this analysis for \SgrA, \VirA, and representative X-ray binaries to a future study.


\begin{acknowledgments}
This work was supported in part by Perimeter Institute for Theoretical Physics and the University of Waterloo.  Research at Perimeter Institute is supported by the Government of Canada through the Department of Innovation, Science and Economic Development Canada and by the Province of Ontario through the Ministry of Economic Development, Job Creation and Trade.
A.E.B. receives additional financial support from the Natural Sciences and Engineering Research Council of Canada through a Discovery Grant. 
\end{acknowledgments}


\appendix

\section{Optical thickness and the transport opacity}
\label{sec:thick}

The relativistic diffusion closure adopted in this work applies when 
the photon mean free path is small in comparison to the scales
on which the thermodynamic variables vary. In a thin, pressure supported settling layer the relevant length is the local (proper) scale height $H$, not the global accretion flow size. Accordingly we take the basic requirement for diffusion to be a transport optical depth across a scale height,

\begin{equation}\label{eq:tauH_def}
\tau_H \,\equiv\, \int \kappa_R \rho\,d\ell \,\sim\, \kappa_R \rho\,H \,\gtrsim\, 1,
\end{equation}
where $\kappa_R$ is the Rosseland mean transport opacity (absorption$+$scattering). This condition is stronger than merely having a large total path length: it guarantees the mean free path is short compared to the gradient scale so that the flux can be written in local diffusive form. If scattering dominates, $\tau_H\gg 1$ ensures trapping and diffusion, while exact LTE/blackbody formation is controlled by an effective absorption depth. 
\subsection{Why a settling layer tends to be optically thick.}

A key point is that an optically thin upstream flow does not imply an optically thin settling layer. In fact, the atmospheric optical depth is controlled by the column accumulated during the time required for the surface to process the supplied baryons. Since $\dot M$ is the mass flux per unit Killing time, the local mass supply per unit proper area and proper time at the foil is
\begin{equation}\label{eq:app:A2}
\dot\Sigma_{\rm loc} \simeq \frac{\dot M}{4\pi r_f^2 N_f}.
\end{equation}
The column contained within one proper scale height is therefore
\begin{equation}\label{eq:Sigma_residence}
\Sigma \sim \dot\Sigma_{\rm loc}\,t_{\rm set}, \qquad
t_{\rm set}\sim\frac{H}{|v_{\hat r,f}|},
\end{equation}
where $v_{\hat r,f}$ is the local radial settling speed measured by a static observer at the base. The kinetic surface condition gives
$|v_{\hat r,f}|\sim |\mathcal{T}|\,c_s(T_f)$, with
$c_s^2=kT/m$, while hydrostatic support gives $H\sim c_s^2/|g_f|$ and
$|g_f|=GM/(N_f r_f^2)$. Combining these relations yields
\begin{align}
\tau_H \sim \kappa_R\Sigma &\sim \frac{\kappa_R\dot M}{4\pi r_f^2N_f}
\frac{H}{|\mathcal{T}|\,c_s(T_f)} \nonumber\\
&\sim \frac{\kappa_R\dot M}{4\pi GM} \frac{c_s(T_f)}{|\mathcal{T}|}.
\label{eq:tauH_scaling}
\end{align}
The lapse cancels between the local mass supply and the proper gravitational acceleration. With $L_{\rm Edd,R}\equiv \frac{4\pi GMc}{\kappa_R}$ and $L_{\rm eq}\equiv \eta\dot M c^2$, the same estimate can be written as
\begin{equation}
\tau_H \sim \frac{L_{\rm eq}}{L_{\rm Edd,R}}\,
\frac{c_s(T_f)}{\eta\,|\mathcal{T}|\,c}.
\end{equation}
Equivalently, in terms of the conventional electron scattering Eddington luminosity $L_{\rm Edd,es}=4\pi GMc/\kappa_{\rm es}$,
\begin{equation}\label{eq:tauH_scaling} 
\tau_H \sim \frac{\kappa_R}{\kappa_{\rm es}}\,
\frac{L_{\rm eq}}{L_{\rm Edd,es}}\,
\frac{c_s(T_f)}{\eta\,|\mathcal{T}|\,c}.
\end{equation}
Thus the diffusion condition $\tau_H\gtrsim1$ requires, parametrically,
\begin{equation}
|\mathcal{T}| \lesssim
\frac{L_{\rm eq}}{L_{\rm Edd,R}}\, \frac{c_s(T_f)}{\eta c}.
\label{eq:T-thick-condition}
\end{equation}
This is a conditional criterion rather than an assertion that every optically thin inflow necessarily produces an optically thick settling layer.

The estimate above addresses the optical depth of the downstream settling layer, not the detailed structure of the inflow-atmosphere transition. In the present framework, the transition radius $r_s$ is treated as either an accretion shock or a smooth matching surface and is closed by the energy and normal stress boundary conditions given in the main text. Indeed, determining $r_s$ and resolving the transition from a particular radiatively inefficient inflow would require a source-dependent radiation MHD calculation. That additional calculation is not required for the local criterion \eqref{eq:T-thick-condition}, but it lies outside the present theory agnostic model.

\medskip

\autoref{eq:tauH_scaling} makes the physics transparent: for fixed accretion supply, the transport depth grows in direct proportion to the time it takes the surface
to process baryons. 
If $\kappa_R=\kappa_{es}$, then for sufficiently small $\mathcal{T}$, $\tau_H\gg1$ and the settling atmosphere is optically thick.

However, the dominant absorption process need not be electron scattering, i.e., $\kappa_R$ can differ from $\kappa_{\rm es}$. 
If absorption in the impinging accretion flow is dominated by synchrotron self-absorption (SSA), as is the case in \SgrA and \VirA, then $\kappa_R$ 
(the relevant Rosseland mean) 
can greatly exceed $\kappa_{\rm es}$ by many orders of magnitude. 

SSA requires a magnetized plasma and a population of relativistic electrons, both of which are naturally supplied by the accretion flow and shock acceleration; upon compression in the surface layer, $n_e$ and $B$ increase, further enhancing the SSA optical depth. 
In our treatment the magnetic field is introduced only as a radiative ingredient (entering $\kappa_R$), and need not be dynamically important.
Nevertheless, if magnetic fields are dynamically important, their added pressure support would tend to increase the scale height/column and thus increase the optical depth.

Even when the upstream flow is optically thin at near-IR frequencies, those that dominate the thermal emission at $T_{\rm eq}$ for \SgrA and \VirA, the accumulated settling layer can become optically thick because synchrotron absorption depends steeply on density.
For a power-law electron distribution $N(\gamma)\propto \gamma^{-p}$ with optically thin index $\alpha=(p-1)/2$, synchrotron self-absorption gives the scaling
\begin{equation}\label{eq-app-alphaB}
\rho\kappa_\nu \equiv \alpha_\nu \,\propto\, n_e\,B^{\alpha+3/2}\,\nu^{-(\alpha+5/2)}.
\end{equation}
With $n_e\propto \rho$ and flux-frozen isotropic compression $B\propto \rho^{2/3}$, this becomes
\begin{equation}\label{eq-app-alpharho}
\alpha_\nu \,\propto\, \rho^{\,2+\frac{2}{3}\alpha}\,\nu^{-(\alpha+5/2)}.
\end{equation}
Thus, while the high frequency optical depth is suppressed relative to mm by the factor $\left(\nu_{\rm NIR}/\nu_{\rm mm}\right)^{-(\alpha+5/2)}$, even a modest density amplification in the settling atmosphere can overwhelm this penalty. For example, taking $\alpha\simeq 1.25$ and $\nu_{\rm NIR}/\nu_{\rm mm}\sim 10^3$ gives a frequency suppression $\sim 10^{-11}$, whereas a density
contrast $\rho_{\rm atm}/\rho_{\rm acc}\sim 10^9$ boosts $\alpha_\nu$ by $\left(10^9\right)^{2+\frac{2}{3}\alpha}\sim 10^{25}$.

Low luminosity black hole systems (e.g., \SgrA and \VirA) are observed to sit near the
synchrotron self-absorption turnover at mm wavelengths. In EHT-motivated one-zone spectral models,
the synchrotron optical depth at 230\,GHz 
are of order unity, with representative values $\tau_{230}\sim 0.4$ for Sgr\,A* and $\tau_{230}\sim 0.2$ for M87*
(depending on model assumptions and position across the image) \cite{M87PaperV,SgrAPaperV}.  Broadband analyses likewise indicate that frequencies below $\sim 230$\,GHz are typically
SSA-thick in the inner M87 system \cite{Algaba_2024}.
Taken together, these constraints motivate using $\tau_{\rm mm}=\mathcal{O}(0.1$-$1)$;
the steep density dependence in \autoref{eq-app-alphaB} and \autoref{eq-app-alpharho} then implies that a
modest increase in column and field strength in the settling layer can drive $\tau_\nu$ to order unity or larger even when the upstream flow is optically thin at higher frequencies.

\subsection{Angle-dependent optical depth.}

 Up to this point we have implicitly characterized the layer by its vertical (radial) optical depth across a scale height, $\tau_\perp\sim\kappa_R\rho H$. However, a geometrically thin layer is strongly anisotropic: its thickness is $\sim H$, while its lateral extent is $\sim R$. Consequently, radiative transfer depends sharply on direction. In a plane parallel approximation, a ray emitted at angle $\theta$ to the local normal traverses a slant column larger by $1/\mu$ with $\mu\equiv\cos\theta$, so that
\begin{equation}
\tau(\mu)\simeq \frac{\tau_\perp}{\mu}.    
\end{equation}
Even when $\tau_\perp\lesssim 1$, a wide range of shallow-angle trajectories can have $\tau(\mu)\gg 1$, leading to repeated scattering/absorption and substantial radiative redistribution along the surface. Therefore, the layer can behave optically thick to much of its own emission even if it is only marginally thick in the vertical direction.

\subsection{Strong field escape geometry.}
The discussion above is purely geometrical and does not invoke gravity. Gravitational lensing can further increase surface self-irradiation by bending trajectories back toward the surface and increasing the number of surface reintersections, but this effect is conceptually distinct: it modifies the path topology, whereas the plane parallel argument above already implies strong angle-dependent transfer in a thin, laterally extended layer.
In fact, near a highly compact surface, non-radial null geodesics reduce the escape solid angle: only a narrow cone of directions reaches future null infinity, while most trajectories re-intersect the atmosphere due to the photon sphere geometry. 

However, gravitational lensing is not expected to alter the local diffusive closure in the optically thick interior.

\subsection{Scope and special cases.}

The diffusion solutions apply to optically thick settling layers with $\tau \equiv \kappa_R P/|g| \gtrsim 1$ and $\kappa_R\simeq\kappa_{\rm es}$. One can, however, envision parameter choices for which the accumulated layer remains optically thin ($\tau\lesssim 1$), e.g., at extremely low $\dot M$, or if the surface processes baryons nearly as fast as they arrive (near-maximal $|\mathcal{T}|\to 1$, preventing
column build-up) then the appropriate closure is optically thin / flux-limited transport rather than diffusion. Likewise, ultra-strong magnetic fields ($B\gtrsim 10^8\,{\rm G}$) that shift cyclotron/synchrotron absorption into the Rosseland band would require frequency dependency transfer and a revised $\kappa_R$.
These regimes lie outside the framework considered here.

\section{Relaxing the slow settling assumption: transonic inner solutions}
\label{app:transonic}

This section extends the steady, spherical atmosphere to allow an inner transonic transition. We retain the relativistic framework, Thomson opacity, and radiative diffusion used in the main text. The only change is that we no longer assume $|u^{r}|\ll c$ at the base: the flow may cross the sonic point once on the way to the foil.  What changes in the transonic case is the treatment of the critical point where the sonic denominator vanishes.

\subsection{Equations of motion}


We keep the same barred variables as in the main text. At the working order of the radiatively dominated closure used in Section \ref{app:sec:relativistic derivation}, the luminosity measured at infinity $L_\infty$ is treated as radially conserved. With gas pressure dominant, we use the same steady, spherically symmetric continuity, Euler, and diffusion equations as in the main text. These reduced equations retain the radial inertial term but were derived under the slow settling and low enthalpy closure. The following calculation is therefore a diagnostic for a possible sonic critical point within the model.

\subsection{Mach number and sonic condition}
The denominator in $\partial_s\bar{x}$, \autoref{eq:R_rEAS}, plays the role of $c_{s}^{2}-v^{2}$. This motivates the (barred) Mach number
\begin{equation}\label{eq:A2_Mach_bar}
\mathcal{M}^{2}(s)\,\equiv\,
\Big(\frac{N}{N_f}\Big)^{\,3}\,
\frac{\mathcal{T}^{2}\,\bar y_f}{\bar x^{2}\,\bar y}\,,
\end{equation}
Therefore
\begin{equation}
    \mathcal{M}=1 \iff \text{sonic}.
\end{equation}
A regular transonic solution that crosses $\mathcal{M}=1$ at the sonic point must satisfy two critical conditions. Let the sonic point be at $r=r_\ast$:
\begin{enumerate}
\item Sonic condition:
\begin{align}
\bar y_\ast \,=\,
\Big(\frac{N_\ast}{N_f}\Big)^{\,3}\,
\frac{\mathcal{T}^{2}\,\bar y_f}{\bar x_\ast^{2}} ,
\label{eq:A2_cond1_bar}
\end{align}
\item Regularity (finite slope, i.e, numerator of \autoref{eq:R_rEAS} vanishes):
\begin{align}
-\frac{N_f}{N_\ast}
+\frac{3\bar y_\ast y_{\rm eq}}{c^2}\frac{N_f}{N_\ast^2}
+\frac{3\tau_hN_f^2L_\infty}{16L_{\rm eq}N_s^2N_\ast\bar y_\ast^3}\bar x_\ast
=0.
\label{eq:A2_cond2_bar}
\end{align}
\end{enumerate}
where $y_{\rm eq}=kT_{\rm eq}/m$, $\tau_h=\kappa\rho_f h$, and $L_{\rm eq}=4\pi r_s^2\sigma T_{\rm eq}^4/N_s^2=\eta\dot{M}c^2$.
Using the temperature equation \autoref{eq:tem_py}, the regularity condition may be written compactly as
\begin{equation}\label{eq:A2_compact_bar}
-\frac{N_f}{N_\ast}
+\frac{3\bar y_\ast y_{\rm eq}}{c^2}\frac{N_f}{N_\ast^2}
-\left(\partial_s\bar y\right)_\ast
=0.
\end{equation}
It means the enthalpy-gravity balance equals the local temperature rise. This gives the finite slopes

\begin{align}
\left.(\partial_s\bar y)\right|_\ast
=
-\frac{N_f}{N_\ast}
+\frac{3N_fy_{\rm eq}\bar y_\ast}{c^2N_\ast^2}
\label{eq:sonic_slope_y},
\end{align}
In Schwarzschild and within the thin layer approximation,
$\partial_s \ln N \simeq (y_{\rm eq}/c^{2})\,N_f/N^{2}$. Define
$$
D(\bar x,\bar y,N)\equiv
\bar y-
\left(\frac{N}{N_f}\right)^3
\frac{\mathcal T^2\bar y_f}{\bar x^2},
$$
and let $F(\bar x,\bar y,N)$ denote the numerator of \autoref{eq:R_rEAS}. At the sonic point, $D_\ast=F_\ast=0$. Since $(\partial_s\bar x)/\bar x=F/D$, the finite density slope follows from l’Hôpital’s rule:

$$
\left.\frac{\partial_s\bar x}{\bar x}\right|_\ast
=
\frac{\left.(dF/ds)\right|_\ast}
{\left.(dD/ds)\right|_\ast}.
$$

The total derivatives are evaluated using the temperature slope above and the Schwarzschild relation for $\partial_sN$. This gives a quadratic equation for $(\partial_s\bar x/\bar x)_\ast$; the physical root is selected by the branch that connects smoothly to the desired solution.

For accretion powered layers one has $L_\infty\le 0$, hence
$\partial_s\bar y>0$. Since $N$ increases outward, \autoref{eq:A2_compact_bar} implies
\begin{equation}\label{eq:b16}
 \frac{3\,\bar y_\ast\,y_{\rm eq}}{c^{2}}\,\frac{N_f}{N_\ast^{2}}
\,>\,\frac{N_f}{N_\ast}
\,\,\Longrightarrow\,\,
3\,\frac{kT_\ast}{m\, c^2}\,>\,1\,.   
\end{equation}
This cleanly explains why transonic solutions are irrelevant in the baryonic temperature regime of interest. Thus a smooth sonic transition would require $kT_\ast>mc^2/3$, i.e. ultra-relativistic baryon temperatures at the sonic point(for protons, $T_\ast\gtrsim 3\times10^{12}\,\mathrm{K}$), well above the regime of interest in the main text. Practically, the solutions remain subsonic; any attempted acceleration would require an inner shock to connect to the solid surface (if a surface is to be reached).


\subsection{Inner boundary and microphysics}

Two remarks are important at the foil $r=r_f$:
\begin{enumerate}[label=\roman*)]
\item Mass flux. In the subsonic model we we imposed a kinetic inflow $J=N_f\,r_f^{2}\,\mathcal{T}\,\rho_f\,\sqrt{kT_f/m}$, \autoref{eq:base_flux_condition}. Here the $N_f$ converts proper to coordinate time and $\mathcal{T}$ parameterizes the wall transmittance relative to a thermal impingement rate.
The sonic point diagnostic above retains this kinetic relation as the mass-flux parameterization used in the reduced equations. It does not specify an independently transonic wall boundary condition. A genuinely transonic or supersonic inner layer would require the unreduced relativistic Euler equation together with a new surface interaction closure, which is outside the scope of the present appendix.


\item Temperature boundary condition. The microphysical base condition from wall-gas interaction (see \autoref{sec:microBC}) is local and unaffected by the Mach number; it only involves the diffusive/conductive flux and the kinetic energy accommodation at the wall. Thus, Equations of microphysics boundary conditions remain unchanged.
\end{enumerate}

\subsection{Summary and Consequences}

Given $(\mathcal{T},\,L_\infty)$, choose a trial sonic location $s_\ast$ and solve the sonic condition \autoref{eq:A2_cond1_bar} together with the regularity
condition \autoref{eq:A2_cond2_bar} for $(\bar x_\ast,\bar y_\ast)$. The corresponding finite derivatives at the sonic point are then obtained from the
temperature equation as written in \autoref{eq:A2_compact_bar} and, equivalently by the the temperature slope relation and the l’Hôpital condition.
These provide a regular eigenslope initializer for integrating the system of \autoref{eq:R_rEAS} and \autoref{eq:tem_py} together, both inward and outward through $r_\ast$.

In the parameter space relevant here, the necessary temperature criterion for a
smooth crossing, $3\,kT_\ast/m > 1$, is not satisfied. Consequently, no global, regular transonic solution exists
under the assumptions of the main text (gas pressure support, gray diffusion, steady, spherical flow).  Practically, allowing a transonic inner region mainly reshapes the density profile near the foil: the atmosphere may accelerate and
thin close to the surface, terminating in a thin shock or boundary layer, but the global energetics are unchanged.

To sum up, (i) the luminosity measured at infinity, $L_\infty$, is radially conserved by the Killing energy current, so the emergent power is set by the accretion power processed through the layer; and (ii) with $L_\infty<0$ the redshifted temperature $T_\infty$ still increases outward, so the layer implying convective stability. In practice, the transonic region is narrow for the cool, optically thick atmospheres considered here; it primarily changes the inner matching to the foil rather than the conserved luminosity result.

\section{Including radiation pressure and radiative inertia at high $L/L_{\rm Edd}$}\label{appsec:radiation_pressure}
This section consider a minimal, self-consistent relativistic upgrade of the settling layer equations that accounts for radiation pressure and radiative inertia when the luminosity approaches the Eddington scale. We keep the steady, spherical, subsonic assumptions and the diffusion closure; the only new ingredients are (i) the radiation contribution to pressure and inertia and (ii) the radiation momentum exchange represented by the radiation pressure gradient in the diffusion limit. The aim is to show where the terms enter and how the equations are modified, without altering the structure of the main text.

\subsection{Total stress energy and basic definitions}

We split the total stress energy into gas and radiation,
\begin{equation}
T^{\mu\nu} \,=\, T^{\mu\nu}_{\rm g} + R^{\mu\nu}.
\end{equation}
For the gas we use the perfect fluid form
$T^{\mu\nu}_{\rm g}=(\varepsilon_{\rm g}+P_{\rm g})u^\mu u^\nu + P_{\rm g}g^{\mu\nu}$ (with $\varepsilon_{\rm g}\simeq \rho c^2$ in our atmosphere).
In the comoving (fluid) frame and in the diffusion limit with Thomson
scattering, the radiation moments are
\begin{equation}
E_{\rm rad}=aT^{4},\quad
P_{\rm rad}=\frac{1}{3}aT^{4},\quad
F^{\mu}.
\end{equation}
Here $F^{\mu}$ is first order in gradients, and $R^{\mu\nu}$ has the usual $(E_{\rm rad},P_{\rm rad},F^\mu)$ structure.

We use the same barred variables as in the main text
\begin{equation}
\bar x(r)\,\equiv\,\frac{N(r)^{2}\,\rho(r)}{N_f^{2}\,\rho_f},
\qquad
\bar y(r)\,\equiv\,\frac{N(r)\,T(r)}{T_{\rm eq}},
\end{equation}
where $y_{\rm eq}$ $y_{\rm eq}$ is given by \autoref{eq:y-equi}, and $N(r)=\sqrt{1-2GM/(rc^{2})}$, and the subscript $f$ denotes the base
($r=r_f$). In terms of $\bar x,\bar y$ the physical fields are
\begin{equation}\label{eq:rho-T-app}
\rho(r) \,=\, \frac{N_f^{2}\rho_f}{N(r)^{2}}\,\bar x(r),
\qquad
T(r) \,=\, \frac{T_{\rm eq}}{N(r)}\,\bar y(r).
\end{equation}
The gas pressure is
\begin{equation}\label{eq:p-gas-app}
P_{\rm g}(r) \,=\, \rho\,\frac{kT}{m}
\,=\, \frac{N_f^{2}\rho_f\,y_{\rm eq}}{N(r)^{3}}\,\bar x(r)\,\bar y(r).
\end{equation}
The radiation pressure is
\begin{equation}\label{eq:p-rad-app}
P_{\rm rad}(r) \,=\, \frac{1}{3}\,a\,T^{4}
\,=\, \frac{a}{3}\,T_{\rm eq}^{4}\,\frac{\bar y(r)^{4}}{N(r)^{4}}.
\end{equation}
(We will also use $E_{\rm rad}=3P_{\rm rad}$ from the radiation moments.) Radiation contributes to the inertia through $(E_{\rm rad}+P_{\rm rad})/c^{2}=4P_{\rm rad}/c^{2}$.
Neglecting the small gas-enthalpy term $P_{\rm g}/c^{2}$ (since $kT\ll mc^{2}$ in the layer), the effective inertia entering the Euler equation is
\begin{align}\label{eq:rho_eff}
\rho_{\rm eff}(r)
&\,\equiv\, \rho(r) \,+\, \frac{4}{c^{2}}\,P_{\rm rad}(r)\nonumber\\
&\,=\, \frac{N_f^{2}\rho_f}{N(r)^{2}}\,\bar x(r)
\,+\, \frac{4a}{3c^{2}}\,T_{\rm eq}^{4}\,\frac{\bar y(r)^{4}}{N(r)^{4}}.
\end{align}
For reference, define the Eddington ratio at infinity
\begin{equation}
\Gamma_\infty \equiv \frac{\kappa L_\infty}{4\pi GMc}.
\label{eq:Gamma_def}
\end{equation}
It is independent of radius because $L_\infty$ is conserved. Since $F_{\hat r}$ below is the flux measured by local static observers, the corresponding local force ratio is $\Gamma_{\rm loc}(r)=\Gamma_\infty/N(r)$, with the sign inherited from $L_\infty$. In the total fluid equation, radiative support is already represented by the radiation pressure gradient; $\Gamma_\infty$ is therefore used only as a diagnostic.


\subsection{ Conservation laws and luminosity}

Total conservation $\nabla_\nu T^{\mu\nu}=0$ still implies a conserved energy
current $J^\mu\equiv -T^{\mu}{}_{\nu}k^{\nu}$ associated with the timelike Killing vector $k^\nu=\partial_t$. As in the main text,
\begin{equation}
L_\infty(r) \,=\, \int_{S_r} J^\mu d\Sigma_\mu,
\end{equation}
is independent of $r$. In other words, the luminosity measured at infinity is the same whether evaluated at the base (from the diffusive flux) or at the top
(from the thermal emission). None of the manipulations below modify this fact.

\subsection{Continuity and diffusion: forms unchanged}

\paragraph{Continuity.}
Spherical steady flow gives
\begin{equation}
\frac{1}{r^{2}}\partial_r(r^{2}\rho u^{r})=0
\quad\Longleftrightarrow\quad
J\equiv -\,r^{2}\rho u^{r}=\mathrm{const}.
\label{eq:continuity_rad}
\end{equation}
identical to the main text.

\paragraph{Diffusion (for $T_\infty$).}
The relativistic diffusion law derived in the main text \autoref{eq:GR-Diff} carries over
\begin{equation}\label{eq:diffusion_rad}
\partial_r T_\infty \,=\,
-\frac{3\,\kappa\,\rho\,N}{64\pi r^{2}\sigma\,T_\infty^{3}}\,L_\infty,
\quad T_\infty\equiv N T.
\end{equation}
Thomson scattering fixes the mean free path $\lambda=(\kappa\rho)^{-1}$, so
radiation pressure/inertia does not alter the transport coefficient itself (if true absorption at the Rosseland peak becomes important, one replaces $\kappa$ by $\kappa_R$ everywhere; the derivation is otherwise identical).


\subsection{Euler with radiation pressure and radiative inertia}
Projecting $\nabla_\nu T^{\mu\nu}=0$ orthogonally to $u^\mu$ yields the radial momentum equation. At the same diffusion order used above, radiation is isotropic in the fluid frame and shares the mean velocity of the gas. Neglecting the small gas-enthalpy contribution $P_g/c^2$, the total-fluid form is
\begin{align}
&\rho_{\rm eff}u^r\partial_r u_r
=
-\partial_r P_{\rm tot}
-\frac{c^2}{2}\rho_{\rm eff}\,\partial_r\ln(N^2),\nonumber\\
&P_{\rm tot}\equiv P_{\rm g}+P_{\rm rad},
\label{eq:Euler_total}
\end{align}
with $\rho_{\rm eff}$ given by \autoref{eq:rho_eff}. For comparison, the gas equation alone may be written as

\begin{align}\label{eq:old32}
&\rho u^r\partial_r u_r
=
-\partial_rP_g
-\frac{\rho c^2}{2}\partial_r\ln(N^2)
+\frac{\kappa\rho}{c}F_r,\nonumber\\
&F_r\equiv \frac{F_{\hat r}}{N},
\end{align}
where $F_{\hat r}$ is the local orthonormal radiative flux. The radiation momentum equation carries the equal and opposite coupling force:

$$
\frac{4P_{\rm rad}}{c^2}u^r\partial_r u_r
=
-\partial_rP_{\rm rad}
-2P_{\rm rad}\,\partial_r\ln(N^2)
-\frac{\kappa\rho}{c}F_r.
$$
Adding this relation to \autoref{eq:old32} gives \autoref{eq:Euler_total}. Thus, the explicit scattering force belongs to the gas equation, whereas the total fluid equation already contains the same momentum exchange through the radiation pressure gradient.

In the static limit $u^r=0$, \autoref{eq:Euler_total} gives

\begin{equation}\label{eq:old33}
\frac{dP_{\rm tot}}{dr}
=-\frac{c^2}{2}\rho_{\rm eff}\frac{d\ln N^2}{dr}.
\end{equation}
In the weak field limit,

\begin{equation}\label{eq:old34}
\frac{d(P_g+P_{\rm rad})}{dr}
=-\left(\rho+\frac{4P_{\rm rad}}{c^2}\right)g
\simeq-\rho g .
\end{equation}
The familiar Eddington reduction appears in the gas only balance,
$$
\frac{dP_g}{dr}
=
-\frac{\rho c^2}{2}\left[1-\Gamma_{\rm loc}(r)\right]
\frac{d\ln N^2}{dr},
$$
which reduces to $dP_g/dr=-\rho g(1-\Gamma_\infty)$ in the Newtonian limit.

Therefore, the Eddington reduction multiplies the gravitational term while radiative inertia drops out of static balance at leading order. The Killing energy
current remains conserved, so the luminosity conservation proof and $L_{\infty}=\text{const}$ are unaffected by the inclusion of radiation pressure and inertia..


\subsection{Sonic discriminant (effective sound speed)}

It is convenient to define
\begin{equation}
\label{eq:chi-bar}
\wp (\bar x,\bar y;N)
\,\equiv\,
\frac{4P_{\rm rad}}{\rho\,c^{2}}\,=\,
\frac{4a}{3c^{2}}\,
\frac{T_{\rm eq}^{4}}{N_f^{2}\rho_f}\,
\frac{\bar y^{4}}{N^{2}\,\bar x},
\end{equation}
and the effective inertia \autoref{eq:rho_eff} becomes
\begin{equation}
\rho_{\rm eff}=\rho\,(1+\wp).
\end{equation}
In the relativistic ODE the sonic denominator \autoref{eq:denominator-px}, comes from the balance of the pressure's response (sound speed piece) against the kinematic term $J^{2}/(N\rho^{2}r^{4})$ generated when eliminating the velocity in favor of $J$. To capture radiative inertia in the same rationalized ODE structure, one may model its effect for example on the sonic discriminant by the substitution $y_{\mathrm{eq}}\bar{y}\mapsto y_{\mathrm{eq}}\bar{y}(1+\wp)$, i.e. by treating $\rho_{\mathrm{eff}}=\rho(1+\wp)$ as an effective enhancement of the pressure response term. Including radiative inertia is equivalent to the replacement
 \begin{equation}
 \underbrace{y_{\rm eq}\,\bar y}_{\text{gas }c_{s}^{2}}
 \quad\,\Longrightarrow\,
 \underbrace{{y_{\rm eq}\,\bar y}\,(1+\wp)}_{\text{gas $+$ radiative inertia}},
 \end{equation}
 hence at the base $r=r_f$ the subsonic base condition \autoref{eq:base-cond-maintext} becomes
 \begin{equation}\label{eq:base-cond-app}
 1+\wp_f-\mathcal{T}^{2}>0 
 \quad\Longleftrightarrow\quad |\mathcal{T}|<\sqrt{1+\, \wp_f},
 \end{equation}
 which reduces to $ |\mathcal{T}|<1 $ when $ \wp_f\ll 1 $. However, note that the radiation coupled critical condition is not obtained by modifying the only gas sonic denominator alone: radiation changes both the pressure response and the inertial term, with the precise relation depending on the adopted radiation closure. We therefore use $\wp$ only as a diagnostic of the importance of radiative inertia. Any coupled radiation critical point must be obtained from the coupled structure equations; no universal replacement of is assumed here.

\subsection{Boundary data and checks}

\paragraph{Top boundary (photosphere / shock interface).}

The normal stress balance across the interface at $r=r_s$ should include radiation pressure on whichever side it is dynamically important. With a subsonic base on the atmospheric side ($u^r(r_s)\simeq 0$), one has
\begin{align}
\label{eq:top-stress}
P_{\rm tot,s}\,\equiv\,P_{{\rm g},s}+P_{{\rm rad},s} &\,=\,
\frac{N_f^{2}\rho_f\,y_{\rm eq}}{N_s^{3}}\,\bar x_s\,\bar y_s
\,+\,\frac{a}{3}\,\frac{T_{\rm eq}^{4}}{N_s^{4}}\,\bar y_s^{4}\nonumber\\
&\,\simeq\,
\underbrace{\frac{\rho_a (u_a^r)^2}{N_s^{2}}+P_a+P_{{\rm rad},a}}_{\text{upstream}},
\end{align}
where the upstream (accretion) side may be simplified to
$P_{{\rm rad},a}\,\approx\,0$, $P_a\,\approx\,0$ for a cold, hypersonic inflow, and $\rho_a u_a^r=\dot M/(4\pi r_s^2)$ by continuity. If upstream dissipation reduces the impact speed, write $u_a^r=\eta_{\rm ram}\,c$ to obtain
\begin{equation}
\label{eq:top-ram-with-rad}
\frac{N_f^{2}\rho_f\,y_{\rm eq}}{N_s^{3}}\,\bar x_s\,\bar y_s
\,+\,\frac{a}{3}\,\frac{T_{\rm eq}^{4}}{N_s^{4}}\,\bar y_s^{4}
\,\simeq\,\eta_{\rm ram}\,\frac{\dot M\,c}{4\pi r_s^{2}\,N_s^{2}}.
\end{equation}
In the regime where $P_{{\rm rad},s}\ll P_{{\rm g},s}$, \autoref{eq:top-ram-with-rad} reduce to the gas-only matching used in the main text (with $\eta_{\rm ram}$
identified with $\eta$ under the same assumptions). The energy balance that sets $L_\infty$ is unchanged
\begin{equation}
\label{eq:Linf-bc}
L_\infty
=\frac{4\pi r_s^2}{N_s^2}\,\sigma\,T_{\infty,s}^{4}-\eta\,\dot M c^{2}= L_{\rm eq}\,(\bar y_s^{4}-1),
\end{equation}
where $T_{\infty,s}=T_{\rm eq}\,\bar y_s$. Since it follows from Killing energy conservation; radiation pressure only enters
through $P_{\rm tot}$ in the stress balance, not in \autoref{eq:Linf-bc}.

\paragraph{Microscopic base boundary.}
The kinetic wall boundary that fixes $T_f$ is local and occurs across a layer of thickness $\lambda_c\,\ll\, h$. Over this microscopic scale the lapse is constant to excellent approximation ($N\simeq N_f$), so the relativistic redshift factors
drop out and the derivation of the temperature floor proceeds exactly as in the main text.
\medskip

\subsection{Validity of the diffusion equation near the photosphere}

In a local orthonormal (hat) frame comoving with the static observers,
let $E$ be the radiation energy density, $F_{\hat r}$ the radial flux,
and $P_{\hat r\hat r}$ the radial radiation pressure. They are the first
angular moments of the specific intensity. Causality bounds $|F_{\hat r}|\le cE$. It is convenient to define the reduced flux
\begin{equation}
f \,\equiv\, \frac{|F_{\hat r}|}{c\,E}\in[0,1] \,,
\qquad
E=aT^{4},\quad a=\frac{4\sigma}{c}.
\end{equation}
The field is nearly isotropic when $f\ll 1$ and free streaming when $f\to 1$.

For Thomson opacity $\kappa$ and mean free path $\lambda=(\kappa\rho)^{-1}$,
the diffusion limit (large optical depth $\tau\equiv \ell/\lambda\gg 1$) gives
\begin{align}
F_{\hat r}\,&\simeq\, -\,\frac{c}{3\kappa\rho}\,\frac{dE}{d\ell}\nonumber\\
&\quad\Rightarrow\quad
f \,=\, \frac{|F_{\hat r}|}{cE} \,\simeq\, \frac{1}{3}\,\frac{\lambda}{\ell}
\,=\, \frac{1}{3\,\tau}\ \ll 1,
\end{align}
where $\ell\equiv E/|dE/d\ell|$ is the local gradient scale. In this regime,
using $E=aT^{4}$ and the GR redshift bookkeeping ($d\ell=dr/N$, $T_\infty\equiv N T$) (\autoref{app:sec:relativistic derivation}) yields the diffusion equation we used in the main text, \autoref{eq:GR-Diff},
\begin{equation}
\partial_r T_\infty
\,=\, -\,\frac{3\,\kappa\,\rho\,N}{64\pi r^{2}\sigma\,T_\infty^{3}}\,L_\infty,
\quad T_\infty\equiv N\,T.
\label{eq:diff-valid}
\end{equation}
If $f$ approaches unity; near the photosphere, the $1/3$ Eddington factor is no longer accurate.
Two standard upgrades preserve causality and smoothly connect diffusion to streaming, without changing the global conservation of Killing energy $L_\infty$.

\begin{enumerate}
\item Flux limited diffusion (FLD).
Keep a diffusion type law but limit the coefficient:
\begin{equation}
F_{\hat r} \,=\, -\,\frac{c\,\lambda_{\rm FL}(R)}{\kappa\rho}\,\frac{dE}{d\ell},
\qquad
R \,\equiv\, \frac{1}{\kappa\rho}\,\frac{|dE/d\ell|}{E},
\end{equation}
with the flux limiter $\lambda_{\rm FL}(R)$ chosen so that
$\lambda_{\rm FL}\,\to\,1/3$ for $R\,\ll\,1$ (diffusion) and
$\lambda_{\rm FL}\,\to\,1/R$ for $R\,\gg\,1$ (free streaming), which enforces
$f=\lambda_{\rm FL}R\le 1$. A common analytic choice is Levermore-Pomraning ( see e.g., \cite{1984JQSRT..31..149L}):
$\lambda_{\rm FL}(R)=\frac{1}{R}\,\left(\coth R-\frac{1}{R}\right)$.
In our spherical, steady setting this amounts to replacing the factor $1/3$ by $\lambda_{\rm FL}$ in the conductivity, i.e.
\begin{equation}
\partial_r T_\infty
= -\,\frac{\kappa\,\rho\,N}{64\pi r^{2}\sigma\,T_\infty^{3}}\,
\frac{1}{\lambda_{\rm FL}(R)}\,L_\infty,
\end{equation}
which reduces to \autoref{eq:GR-Diff} when $\lambda_{\rm FL}=1/3$.

\item two-moment closure (M1).
Evolve two moments $(E,F_{\hat r})$ and close the pressure tensor
with an Eddington factor that depends on $f$
\begin{equation}
P_{\hat r\hat r}=\chi(f)\,E,\quad
\chi(f)=\frac{3+4f^{2}}{5+2\sqrt{\,4-3f^{2}\,}},
\end{equation}
so $\chi\,\to\,1/3$ as $f\,\to\,0$ and
$\chi\,\to\,1$ as $f\,\to\,1$. In curved spacetime all relations are evaluated in the local orthonormal frame; $F_{\hat r}$ is tied to $L_\infty$ by
$L_\infty=4\pi r^{2}N^2\,F_{\hat r}$, and the moment equations couple to the gas through $\kappa$ exactly as in diffusion.
\end{enumerate}
Both FLD and M1 change only the constitutive relation between the local
radiation flux/pressure and gradients. The global conserved current
$J^\mu\equiv -\,T^{\mu}{}_{\nu}\,k^\nu$ (with $T^{\mu\nu}$ the total
gas$+$radiation stress energy and $k^\nu=\partial_t$ the Killing vector) still
obeys $\nabla_\mu J^\mu=0$ in steady state. Hence
\begin{equation}
L_\infty(r)=\int_{S_r} J^\mu d\Sigma_\mu,
\end{equation}
is radius-independent irrespective of whether one uses pure diffusion,
FLD, or M1 in the thin transition layer.

\subsection{Summary and Consequences}
To include radiation pressure and radiative inertia in practice, use the total fluid Euler equation, \autoref{eq:Euler_total}, with $P_{\rm tot}=P_g+P_{\rm rad}$ and $\rho_{\rm eff}=\rho+4P_{\rm rad}/c^2$. Radiation support is then represented by $\partial_rP_{\rm rad}$. Equivalently, one may use the gas equation, \autoref{eq:old32}, together with the corresponding radiation momentum balance. The continuity law and the diffusion equation retain their forms in the diffusion regime, and $P_{\rm rad}$ should be included in the normal stress balance when dynamically relevant.


\medskip

To sum up, radiation affects the settling layer in two ways: 
(i) it reduces the effective weight by the Eddington factor $g\to g_{\rm eff}\,\equiv g(1-\Gamma)$ with $\Gamma=\kappa L_\infty/(4\pi G\,Mc)$ in our model; and (ii) it adds radiative inertia to the sonic denominator which we modeled in the rationalized ODE by an effective modification of the sonic discriminant controlled by $\rho_{\rm eff}=\rho(1+\wp)$ with $\wp\equiv 4P_{\rm rad}/(\rho c^{2})>0$. Thus, the sound speed term is multiplied by $(1+\wp)$, which pushes any transonic point inward for fixed thermodynamic state. As $\Gamma\to 1$ hydrostatic balance flattens ($g_{\rm eff}\to 0$), the density gradient can weaken, and the layer can become radiation pressure dominated; nevertheless the Killing energy current remains conserved, so the luminosity at infinity is still radius independent: $L_\infty(r)=\mathrm{const}$. Thought, the parameter $\wp=4P_{\rm rad}/(\rho c^2)$ measures the importance of radiative inertia, a radiation coupled critical condition must be obtained from the coupled closure. The diffusion equation \autoref{eq:diffusion_rad} remains valid while $f\ll1$; near the photosphere, where $f$ approaches unity, a flux-limited (FLD) r two-moment closure M1 should replace the $1/3$ diffusion closure locally without changing the global Killing energy conservation.


\section{Classical derivation of radiative diffusion equation}\label{appsec: classic_rad}
We begin by noting that the temperature as measured at infinity (e.g., a color temperature) is related to that in the comoving frame by the gravitational redshift, $T_\infty=N T$.  The luminosity of a thermally emitting spherical surface with radius $r$ as measured at infinity is given by
\begin{equation}\label{eq:lum_infty}
    L_\infty(r,T_\infty) = A_\infty(r) \sigma T_\infty^4 = \frac{4\pi r^2}{N^2} \sigma T_\infty^4,
\end{equation}
Here $A_\infty(r)=4\pi r^2/N(r)^2$ is the apparent emitting area for a static, spherical surface outside the photon sphere, $r\geq 3r_g$. For $2r_g<r<3r_g$, the escape cone is restricted by photon capture, and the observed area must be obtained from the corresponding escape-cone (or ray-tracing) relation rather than the simple $N^{-2}$ scaling. That is why the restriction exists.

Also, $T_\infty$ and $L_\infty$ provide convenient, gauge-invariant definitions of temperature and luminosity in terms that are observationally relevant.

If the emission region is not moving (i.e., $u^r=0$), the local emitted flux is $F=\sigma T^4 = \sigma (T_\infty/N)^4$, where the factor of $N$ accounts for the gravitational redshift.  Thus, the flux from a spherical surface at radius $r$ as measured at a location $r'$ is
\begin{equation}
    F(r'; r) = \sigma \left[ T_\infty(r)/N(r') \right]^4.
\end{equation}
Therefore, the net flux through a surface at radius $r$ between two
spherical surfaces located at $r-\Delta r/2$ and $r+\Delta r/2$ is
\begin{equation}
\begin{aligned}
F_{\rm net}(r)
&=F(r;r-\Delta r/2)-F(r;r+\Delta r/2)\\
&=\frac{\sigma}{N(r)^{4}}
\Big[T_\infty(r-\Delta r/2)^{4}-T_\infty(r+\Delta r/2)^{4}\Big]\\
&\approx
-\frac{4\sigma}{N^{4}}\,T_\infty^{3}\,\partial_r T_\infty\,\Delta r,
\end{aligned}
\end{equation}
where $N$ and $T_\infty$ are evaluated at $r$.

Within the diffusion approximation, photons random-walk a characteristic
proper distance $\ell\simeq \lambda/3$ per step along the radial direction,
where $\lambda=(\kappa\rho)^{-1}$ is the comoving mean free path and the
factor $1/3$ is the usual angular average.  In Schwarzschild,
the proper radial distance is $d\ell=dr/N$, so over one step
\begin{equation}
\Delta r \simeq N\,\ell \simeq \frac{N\lambda}{3}.
\end{equation}
Substituting this into the expression above gives
\begin{equation}
\partial_r T_\infty
=
-\frac{3\kappa\rho}{N}\,
\frac{N^{4}}{16\sigma T_\infty^{3}}\,
F_{\rm net}(r).
\end{equation}
Finally, using $A_\infty(r)=4\pi r^2/N^2$ and $L_\infty=A_\infty\sigma T_\infty^4$,
the net local flux is related to $L_\infty$ by
\begin{equation}
N^4 F_{\rm net}(r)=\frac{N^2}{4\pi r^2}\,L_\infty .
\end{equation}
 we have
\begin{equation} \label{eq:R_RadDiff}
\partial_r T_\infty = -\frac{3\kappa\rho N L_\infty}{64\pi r^2 \sigma T_\infty^3}.
\end{equation}
When expressed this way, the radiative diffusion equation matches \autoref{eq:GR-Diff}, with the redefinition of $T$ to be $T_\infty$ and $L$ to be $L_\infty$, thereby incorporating the gravitational redshift and lensing naturally.  We note that when $L_\infty=0$, i.e., the source is in global thermal equilibrium, $T_\infty$ is constant.

\section{Time dependence and validity of the steady approximation}
\label{app:timedep}

The main text constructs steady ($\partial_t=0$), spherically symmetric settling layers.
Here we investigate the corresponding time dependent conservation statements and the resulting conditions under which the steady solutions may interpret as quasi-static.
This section is intentionally general, meaning it does not solve the time dependent radiation-hydrodynamic problem, but it makes explicit what changes when $\partial_t\neq 0$.

\subsection{Killing energy conservation with $\partial_t\neq 0$}
\label{app:timedep_killing}

Let $K^\mu=(\partial/\partial t)^\mu$ be the timelike Killing vector of the exterior
Schwarzschild spacetime, and define the associated conserved current
\begin{equation}
J^\mu \;\equiv\; T^{\mu\nu} K_\nu,
\qquad \nabla_\mu J^\mu = 0,
\end{equation}
where $T^{\mu\nu}$ is the total stress energy tensor. Indeed, no stationarity assumption is required for $\nabla_\mu J^\mu=0$; it follows from $\nabla_\mu T^{\mu\nu}=0$ and Killing's equation for $K^\mu$. In this appendix, $L_\infty$ denotes the exact luminosity associated with the complete Killing energy current. In the steady atmospheric calculation of \autoref{app:sec:relativistic derivation}, we retain its leading diffusive contribution under the radiatively dominated closure stated following Equation (48).

Now considering the spacetime 4-volume $\mathcal{V}$ bounded by two constant-$t$ hypersurfaces $\Sigma_t$ and $\Sigma_{t+\Delta t}$ and by two timelike worldtubes at fixed areal radii $r=r_1$ and $r=r_2$ ($r_2>r_1$). Gauss theorem gives us
\begin{equation}
\int_{\mathcal{V}} \nabla_\mu J^\mu \sqrt{-g}\,d^4x \;=\;
\int_{\partial\mathcal{V}} J^\mu\,d\Sigma_\mu \;=\;0.
\end{equation}
Writing the boundary integral as contributions from the time slices and the radial worldtubes yields the exact energy-balance identity
\begin{align}\label{eq:timedep_energy_balance}
E_\infty(r_1,r_2;t&+\Delta t)-E_\infty(r_1,r_2;t)\\[6pt]
&\;=\; \int_t^{t+\Delta t} \,\, dt'\,\Big[\,L_\infty(r_1,t')- L_\infty(r_2,t')\,\Big],\nonumber
\end{align}
where the Killing energy contained between $r_1$ and $r_2$ at time $t$ is given by
\begin{align}
E_\infty(r_1,r_2;t) &\;\equiv\; -\,\int_{\Sigma_t\cap\{r_1<r<r_2\}} J^\mu n_\mu \, d\Sigma \\[5pt]
&\;=\; 4\pi \int_{r_1}^{r_2} r^2\,J^t(r,t)\,dr,\nonumber
\label{eq:Einf_def}
\end{align}
also, the (outward) luminosity measured at infinity through a sphere of areal radius $r$ is the Killing energy flux per unit Killing time,
\begin{align}
L_\infty(r,t) & \;\equiv\; -\,\int_{S_r} J^\mu\,d\Sigma_\mu\Big/ dt \\[4pt]
&\;=\; -\,\int_{S_r} J^\mu\,n_\mu\,N\,dA \;=\; -\,4\pi r^2\,J^r(r,t).\nonumber
\label{eq:Linf_timedep_def}
\end{align}
Here $S_r$ is the $2$-sphere of areal radius $r$ at fixed $t$ with proper area element
$dA=r^2\sin\theta\,d\theta\,d\phi$. Also $n_\mu=(0,N^{-1},0,0)$ is the outward unit normal to the worldtube $r=\mathrm{const}$, whose induced determinant satisfies $\sqrt{-h}=Nr^2\sin\theta$ so that $d\Sigma_\mu=n_\mu\,\sqrt{-h}\,dt\,d\theta d\phi=n_\mu N\,dA\,dt$.

Now taking $\Delta t\to 0$ in \eqref{eq:timedep_energy_balance} gives the differential form
\begin{equation}
\frac{d}{dt}E_\infty(r_1,r_2;t)\;=\; L_\infty(r_1,t)-L_\infty(r_2,t).
\label{eq:timedep_energy_balance_diff}
\end{equation}
Equation \eqref{eq:timedep_energy_balance_diff} then makes two key points precise, as follows:

\begin{enumerate}
\item In steady state ($dE_\infty/dt=0$), the luminosity is radially constant:
$L_\infty(r,t)=\mathrm{const}$, which is the statement used throughout the main text.

\item In non-steady situations ($dE_\infty/dt\neq 0$), the luminosity generally "is not"
radially constant: differences in $L_\infty$ between radii are exactly accounted for by the rate of change of stored Killing energy in the intervening region.
\end{enumerate}
For the settling layer, taking $(r_1,r_2)=(r_f,r_s)$ shows that the emergent luminosity at the photosphere/shock satisfies
\begin{equation}
L_\infty(r_s,t) \;=\; L_\infty(r_f,t) \;-\;\frac{d}{dt}E_\infty(r_f,r_s;t),
\label{eq:Lout_Lin_storage}
\end{equation}
where $E_\infty(r_f,r_s;t)$ is the total Killing energy contained in the atmospheric layer. Thus, when the injected power increases, part of it goes into increasing $E_\infty$ and the emergent luminosity rises more slowly; when the injected power decreases, the layer can release stored energy ($dE_\infty/dt<0$) and the emergent luminosity remains elevated for a time. In this sense, time variability is naturally interpreted as buffering by atmosphere, i.e., the atmosphere acts as an energy reservoir that smooths and time delays variability.
\subsection{Response timescales and quasi-static regime}
\label{app:timedep_timescales}
The steady solutions are expected to hold quasi-statically when the accretion supply varies on a timescale long enough compared to the layer response times. Two immidiate estimates are as follows.

\begin{enumerate}
    \item Settling time, $t_{\rm set}$: the time for baryons to drift through one scale height is
\begin{equation}\label{eq:tset_def}
t_{\rm set}\sim \frac{H}{|v_r|},
\end{equation}
which may also be written in terms of the column $\Sigma\sim \rho H$ as
$t_{\rm set}\sim 4\pi r_f^2\,\Sigma/\dot M$. With the kinetic closure at the base $|v_r|_f \sim |\mathcal{T}|\,c_s(T_f)$ and hydrostatic support
$H\sim c_s^2/|g|$, one finds the scaling $t_{\rm set}\sim c_s/(|\mathcal{T}|\,|g|)$ (up to order unity factors).

\item Radiative diffusion time, $t_{\rm diff}$: in diffusion, the characteristic escape time across a scale height is
\begin{equation}
t_{\rm diff}\sim \frac{H^2}{D} \sim \frac{3\kappa_R\rho\,H^2}{c}
\sim 3\,\tau_H\,\frac{H}{c},
\label{eq:tdiff_def}
\end{equation}
where $\tau_H\equiv \int \kappa_R\rho\,d\ell \sim \kappa_R\rho H$, and $D\sim c/(3\kappa_R\rho)$ is the diffusion coefficient, $d\ell=dr/N$ is proper radial distance, and $\kappa_R$ is the Rosseland mean transport opacity.
(note that only the scaling is needed here; the numerical prefactor is not important for our purposes.)
\end{enumerate}
\paragraph{Formation of the settling column.} The steady column is an asymptotic state. Let $\dot{\Sigma}_{\rm loc}$ denote the local mass supply rate per unit area, given by \autoref{eq:app:A2}. At fixed supply and base conditions, a one-zone estimate for the buildup of a characteristic scale height column is

\begin{equation}
\frac{d\Sigma}{dt}\simeq \dot{\Sigma}_{\rm loc}-\frac{\Sigma}{t_{\rm set}},
\qquad
\Sigma(t)\simeq \Sigma_{\rm ss}\left(1-e^{-t/t_{\rm set}}\right),
\end{equation}
where $\Sigma_{\rm ss}\simeq \dot{\Sigma}_{\rm loc}t_{\rm set}$. Accordingly,
\begin{align}
&\tau_H(t)\simeq \tau_{H,{\rm ss}}\left(1-e^{-t/t_{\rm set}}\right),\nonumber\\
&t_{\tau=1}\simeq -t_{\rm set}\ln\,\left(1-\tau_{H,{\rm ss}}^{-1}\right).
\end{align}
When $\tau_{H,{\rm ss}}\gg1$, this gives $t_{\tau=1}\simeq t_{\rm set}/\tau_{H,{\rm ss}}\simeq(\kappa_R\dot{\Sigma}_{\rm loc})^{-1}$. Thus, in the leading fixed thermodynamics scaling, the time to become optically thick need not grow as $\mathcal T$ decreases, even though the time to establish the full stationary column does. The steady solution is applicable only when the supply remains approximately unchanged for
\begin{equation}
t_{\rm avail}\gg t_{\rm form}\equiv
\max\,\left(t_{\rm set},t_{\rm diff,ss}\right).
\end{equation}
For $0<\mathcal T<1$, the steady scalings give $t_{\rm set}\propto\mathcal T^{-1}$ and $\tau_{H,{\rm ss}}\propto\mathcal T^{-1}$. Hence very small $\mathcal T$ produces a denser, more optically thick layer, but also requires a correspondingly longer-lived supply to establish the complete steady profile. At $\mathcal T=0$, $t_{\rm set}$ diverges and $\Sigma\simeq\dot{\Sigma}_{\rm loc}t$; this limit is time dependent and lies outside the present steady treatment.

\paragraph{Quasi-static criterion}
Let $t_{\rm var}$ denote the characteristic timescale on which the outer supply parameters vary (e.g.\ $\dot M(t)$ and the associated injected power). If
\begin{equation}
t_{\rm var}\gg t_{\rm resp} \;\equiv\; \max\,(t_{\rm set},t_{\rm diff}),
\label{eq:quasistatic_condition}
\end{equation}
then the atmosphere evolves in a quasi-static way and the steady profiles may be interpreted as a sequence of instantaneous steady states with slowly varying parameters.
If instead $t_{\rm var}\lesssim t_{\rm diff}$, the emergent luminosity cannot track the injected power instantaneously. \autoref{eq:Lout_Lin_storage} then implies that the mismatch is controlled by the time derivative of the stored energy $E_\infty(r_f,r_s;t)$, i.e., variability produces temporal phase rather than a persistent suppression of the time-averaged luminosity.

\subsection{Implications for hiding accretion power}
\label{app:timedep_implications}

Time dependence modifies the instantaneous relation between injected power and emergent
luminosity through the storage term in \eqref{eq:Lout_Lin_storage}. However, it does not provide a generic mechanism to suppress the luminosity on long timescales
compared to $t_{\rm resp}$: averaging \eqref{eq:timedep_energy_balance_diff} over a time interval $\Delta t\gg t_{\rm resp}$ gives
\begin{align}
\langle L_\infty(r_2)\rangle&=\langle L_\infty(r_1)\rangle
-\frac{E_\infty(t+\Delta t)-E_\infty(t)}{\Delta t}\nonumber\\
\,&\;\simeq\; \langle L_\infty(r_1)\rangle,
\end{align}
since $E_\infty$ cannot grow without bound in a finite mass layer. Therefore, within the assumptions of locality and opacity confined transport, the steady state luminosity
result should be interpreted as applying to time averaged emission whenever the supply persists for times longer than the layer response time.

Of course, a complete treatment of strongly time dependent accretion would require solving the coupled time dependent radiation in hydrodynamic system and specifying additional microphysics at the outer interface; this lies beyond the scope of the present paper and motivates future work.

5. Very small transmission and steady-state formation

The mechanism becomes stronger as $\mathcal T$ decreases:

$$ \rho_f\propto\mathcal T^{-1}, \qquad \tau\propto\mathcal T^{-1}. $$

But the processing or settling time also grows:

$$ t_{\rm set}\propto\mathcal T^{-1}. $$

Therefore the paper must show that an optically thick steady layer forms within the source lifetime.

At $\mathcal T=0$, there is no steady transmitted mass flux; the atmosphere continually accumulates matter. The manuscript should consistently use $0<\mathcal T<1$ and treat $\mathcal T=0$ as time dependent.

\bibliographystyle{aasjournal_aeb}
\bibliography{references,EHTCPapers}

\end{document}